\documentclass[useAMS,usegraphicx,usenatbib]{mn2e}          

\usepackage{times}

\newcommand{\Msun}{\ensuremath{\,{\rm M}_\odot}}                  
\newcommand{\Rsun}{\ensuremath{\,{\rm R}_\odot}}                  
\newcommand{\Teff}{\ensuremath{T_{\rm eff}}}                      
\newcommand{\logg}{\ensuremath{\log g}}                           
\newcommand{\Mjup}{\ensuremath{\,{\rm M}_{\rm Jup}}}              
\newcommand{\Rjup}{\ensuremath{\,{\rm R}_{\rm Jup}}}              
\newcommand{\Teq}{\ensuremath{T_{\rm eq}^{\,\prime}}}             
\newcommand{\safronov}{\ensuremath{\Theta}}                       
\newcommand{\ms}{\,m\,s$^{-1}$}                                   
\newcommand{\mss}{\,m\,s$^{-2}$}                                  
\newcommand{\as}{\ensuremath{^{\prime\prime}}}                    
\newcommand{\am}{\ensuremath{^\prime}}                            
\newcommand{\FeH}{\ensuremath{\left[\frac{\rm Fe}{\rm H}\right]}} 
\newcommand{\pjup}{\ensuremath{\,\rho_{\rm Jup}}}                 
\newcommand{\psun}{\ensuremath{\,\rho_\odot}}                     
\newcommand{\chir}{\ensuremath{\chi_\nu^{\,2}}}                   
\newcommand{\mc}[1]{\multicolumn{2}{c}{#1}}
\newcommand{\mcc}[1]{\multicolumn{3}{c}{#1}}
\newcommand{\er}[3]{\ensuremath{#1^{+#2}_{-#3}}}

\newcommand{\ermcc}[5]{\mcc{\ensuremath{{#1\,^{+#2}_{-#3}}\,^{+#4}_{-#5}}}}
\newcommand{\reff}[1]{{#1}}

\title[WASP-24, WASP-25 and WASP-26]
      {High-precision photometry by telescope defocussing. VI. WASP-24, WASP-25 and WASP-26%
      \thanks{Based on data collected by MiNDSTEp with the Danish 1.54\,m telescope at the ESO La Silla Observatory.}}

\author[Southworth et al.]
       {John Southworth\,$^{1}$,
        T.\ C.\ Hinse\,$^{2}$,
        M.\ Burgdorf\,$^{3}$,
        S.\ Calchi Novati\,$^{4,5}$,
        M.\ Dominik\,$^{6}$\thanks{Royal Society University Research Fellow}
        \newauthor
        P.\ Galianni\,$^{6}$,
        T.\ Gerner\,$^{7}$,
        E.\,Giannini\,$^{7}$,
        S.-H.\ Gu\,$^{8,9}$,
        M.\ Hundertmark\,$^{6}$,
        U.\ G.\ J{\o}rgensen\,$^{10}$,
        \newauthor
        D.\ Juncher\,$^{10}$,
        E.\ Kerins\,$^{11}$,
        L.\ Mancini\,$^{12}$,
        M.\ Rabus\,$^{13,12}$,
        D.\ Ricci\,$^{14}$,
        S.\ Sch\"afer\,$^{15}$,
        \newauthor
        J.\ Skottfelt\,$^{10}$,
        J.\ Tregloan-Reed\,$^{1,16}$,
        X.-B.\ Wang\,$^{8,9}$,
        O.\ Wertz\,$^{17}$,
        K.\ A.\ Alsubai\,$^{18}$,
        \newauthor
        J.\ M.\ Andersen\,$^{10,19}$,
        V.\ Bozza\,$^{4,20}$,
        D.\ M.\ Bramich\,$^{18}$,
        P.\ Browne\,$^{6}$,
        S.\ Ciceri\,$^{12}$,
        \newauthor
        G.\ D'Ago\,$^{4,20}$,
        Y.\ Damerdji\,$^{17}$,
        C.\ Diehl\,$^{7,21}$,
        P.\ Dodds\,$^{6}$,
        A.\,Elyiv\,$^{22,17,23}$,
        X.-S.\ Fang\,$^{8,9}$,
        \newauthor
        F.\ Finet\,$^{17,24}$,
        R.\ Figuera Jaimes$^{6,25}$,
        S.\ Hardis\,$^{10}$,
        K.\ Harps{\o}e$\,^{10}$,
        J.\ Jessen-Hansen\,$^{26}$,
        \newauthor
        N.\ Kains\,$^{27}$,
        H.\ Kjeldsen\,$^{26}$,
        H.\ Korhonen\,$^{28,10}$,
        C.\ Liebig\,$^{6}$,
        M.\ N.\ Lund\,$^{26}$,
        M.\ Lundkvist\,$^{26}$,
        \newauthor
        M.\ Mathiasen\,$^{10}$,
        M.\ T.\ Penny\,$^{29}$,
        A.\ Popovas\,$^{10}$,
        S.\ Proft\,$^{7}$,
        S.\ Rahvar\,$^{30}$,
        K.\ Sahu\,$^{27}$,
        \newauthor
        G.\ Scarpetta\,$^{5,4,20}$,
        R.\ W.\ Schmidt\,$^{7}$,
        F.\ Sch\"onebeck\,$^{7}$,
        C.\ Snodgrass\,$^{31}$,
        R.\ A.\ Street\,$^{32}$,
        \newauthor
        J.\ Surdej\,$^{17}$,
        Y.\ Tsapras\,$^{32,33}$,
        C.\ Vilela\,$^{1}$,
        \\
        $^{~~1}$\,Astrophysics Group, Keele University, Staffordshire, ST5 5BG, UK \\
        $^{~~2}$\,Korea Astronomy and Space Science Institute, Daejeon 305-348, Republic of Korea \\
        $^{~~3}$\,HE Space Operations GmbH, Flughafenallee 24, D-28199 Bremen, Germany \\
        $^{~~4}$\,Dipartimento di Fisica ``E.R. Caianiello'', Universit\`a di Salerno, Via Giovanni Paolo II 132, 84084, Fisciano (SA), Italy \\
        $^{~~5}$\,Istituto Internazionale per gli Alti Studi Scientifici (IIASS), 84019 Vietri Sul Mare (SA), Italy \\
        $^{~~6}$\,SUPA, University of St Andrews, School of Physics \& Astronomy, North Haugh, St Andrews, KY16 9SS, UK \\
        $^{~~7}$\,Astronomisches Rechen-Institut, Zentrum f\"ur Astronomie, Universit\"at Heidelberg, M\"onchhofstra{\ss}e 12-14, 69120 Heidelberg, Germany \\
        $^{~~8}$\,Yunnan Observatories, Chinese Academy of Sciences, Kunming 650011, China \\
        $^{~~9}$\,Key Laboratory for the Structure and Evolution of Celestial Objects, Chinese Academy of Sciences, Kunming 650011, China \\
        $^{10}$\,Niels Bohr Institute \& Centre for Star and Planet Formation, University of Copenhagen, Juliane Maries vej 30, 2100 Copenhagen \O, Denmark \\
        $^{11}$\,Jodrell Bank Centre for Astrophysics, University of Manchester, Oxford Road, Manchester M13 9PL, UK \\
        $^{12}$\,Max Planck Institute for Astronomy, K\"onigstuhl 17, 69117 Heidelberg, Germany \\
        $^{13}$\,Instituto de Astrof\'isica, Facultad de F\'isica, Pontificia Universidad Cat\'olica de Chile, Av.\ Vicu\~na Mackenna 4860, 7820436 Macul, Santiago, Chile \\
        $^{14}$\,Instituto de Astronom\'{\i}a -- UNAM, Km 103 Carretera Tijuana Ensenada 422860, Ensenada (Baja Cfa), Mexico \\
        $^{15}$\,Institut f\"ur Astrophysik, Georg-August-Universit\"at G\"ottingen, Friedrich-Hund-Platz 1, 37077 G\"ottingen, Germany \\
        $^{16}$\,NASA Ames Research Center, Moffett Field, CA, USA \\
        $^{17}$\,Institut d'Astrophysique et de G\'eophysique, Universit\'e de Li\`ege, 4000 Li\`ege, Belgium \\
        $^{18}$\,Qatar Environment and Energy Research Institute, Qatar Foundation, Tornado Tower, Floor 19, P.O.\ Box 5825, Doha, Qatar \\
        $^{19}$\,Department of Astronomy, Boston University, 725 Commonwealth Avenue, Boston, MA 02215, USA \\
        $^{20}$\,Istituto Nazionale di Fisica Nucleare, Sezione di Napoli, Napoli, Italy \\
        $^{21}$\,Hamburger Sternwarte, Universit\"at Hamburg, Gojenbergsweg 112, 21029 Hamburg, Germany \\
        $^{22}$\,Dipartimento di Fisica e Astronomia, Universit\`a di Bologna, Viale Berti Pichat 6/2, I-40127  Bologna, Italy \\
        $^{23}$\,Main Astronomical Observatory, Academy of Sciences of Ukraine, vul. Akademika Zabolotnoho 27, 03680 Kyiv, Ukraine \\
        $^{24}$\,Aryabhatta Research Institute of Observational Sciences (ARIES), Manora Peak, Nainital-263 129, Uttarakhand, India \\
        $^{25}$\,European Southern Observatory, Karl-Schwarzschild-Stra{\ss}e 2, 85748 Garching bei M\"unchen, Germany \\
        $^{26}$\,Stellar Astrophysics Centre (SAC), Department of Physics and Astronomy, Aarhus University, Ny Munkegade 120, DK-8000 Aarhus C, Denmark \\
        $^{27}$\,Space Telescope Science Institute, 3700 San Martin Drive, Baltimore, MD 21218, USA \\
        $^{28}$\,Finnish Centre for Astronomy with ESO (FINCA), University of Turku, V{\"a}is{\"a}l{\"a}ntie 20, FI-21500 Piikki{\"o}, Finland \\
        $^{29}$\,Department of Astronomy, Ohio State University, 140 W.\ 18th Ave., Columbus, OH 43210, USA \\
        $^{30}$\,Department of Physics, Sharif University of Technology, P.\,O.\,Box 11155-9161 Tehran, Iran \\
        $^{31}$\,Max Planck Institute for Solar System Research, Justus-von-Liebig-Weg 3, 37077 G\"ottingen, Germany \\
        $^{32}$\,LCOGT, 6740 Cortona Drive, Suite 102, Goleta, CA 93117, USA \\
        $^{33}$\,School of Mathematical Sciences, Queen Mary, University of London, Mile End Road, London E1 4NS, UK \\
        }

\begin{document} \maketitle 

\clearpage

\begin{abstract}
We present time-series photometric observations of thirteen transits in the planetary systems WASP-24, WASP-25 and WASP-26. All three systems have orbital obliquity measurements, WASP-24 and WASP-26 have been observed with {\it Spitzer}, and WASP-25 was previously comparatively neglected. Our light curves were obtained using the telescope-defocussing method and have scatters of 0.5 to 1.2 mmag relative to their best-fitting geometric models. We used these data to measure the physical properties and orbital ephemerides of the systems to high precision, finding that our improved measurements are in good agreement with previous studies. High-resolution {\it Lucky Imaging} observations \reff{of all three targets} show no evidence for faint stars close enough to contaminate our photometry. We confirm the eclipsing nature of the star closest to WASP-24 and present the detection of a detached eclipsing binary within 4.25\,arcmin of WASP-26.
\end{abstract}



\section{Introduction}                                                                                                              \label{sec:intro}

Whilst there are over 1000 extrasolar planets now known, much of our understanding of these objects rests on those which transit their parent star. For these exoplanets only is it possible to measure their radius and true mass, allowing the determination of their surface gravity and density, and thus inference of their internal structure and formation processes.

A total of 1137\footnote{Data taken from the Transiting Extrasolar Planet Catalogue (TEPCat) available at: {\tt http://www.astro.keele.ac.uk/jkt/tepcat/} on the date 2014/07/16.} transiting extrasolar planets (TEPs) are now known, but only a small fraction of these have high-precision measurements of their physical properties. Of these 1150 planets, 58 have mass and radius measurements to 5\% precision, and only eight to 3\% precision.

The two main limitations to the high-fidelity measurements of the masses and radii of TEPs are the precision of spectroscopic radial velocity measurements (mainly affecting objects discovered using the CoRoT and {\it Kepler} satellites) and the quality of the transit light curves (for objects discovered via ground-based facilities). Whilst the former problem is intractable with current instrumentation, the latter problem can be solved by obtaining high-precision transit light curves of TEP systems which are bright enough for high-precision spectroscopic observations to be available.

We are therefore undertaking a project to characterise bright TEPs visible from the Southern hemisphere, using the 1.54\,m Danish Telescope in defocussed mode. In this work we present transit light curves of three targets discovered by the SuperWASP project \citep{Pollacco+06pasp}. From these, and published spectroscopic analyses, we measure their physical properties and orbital ephemerides to high precision.

\subsection{WASP-24}                                                                                                        \label{sec:intro:w24}

This planetary system was discovered by \citet{Street+10apj} and consists of a Jupiter-like planet (mass 1.2\Mjup\ and radius 1.3\Rjup) on a circular orbit around a late-F star (mass 1.2\Msun\ and radius 1.3\Rsun) every 2.34\,d. The comparatively short orbital period and hot host star means WASP-24\,b has a high equilibrium temperature of 1800\,K. \citet{Street+10apj} obtained photometry of eight transits, of which three were fully observed, from the Liverpool, Faulkes North and Faulkes South telescopes (LT, FTN and FTS). The nearest star to WASP-24 (21.2\as) was found to be an eclipsing binary system with 0.8\,mag deep eclipses on a possible period of 1.156\,d.

\citet{Simpson+11mn} obtained high-precision radial velocities (RVs) of one transit using the HARPS spectrograph. From modelling of the Rossiter-McLauglin (RM) effect \citep{Rossiter24apj,McLaughlin24apj} they found a projected spin-orbit alignment angle $\lambda = -4.7 \pm 4.0^\circ$. This is consistent with WASP-24\,b having zero orbital obliquity.

\citet{Smith+12aa} presented observations of two occultations (at 3.6\,$\mu$m and 4.5\,$\mu$m) with the {\it Spitzer} space telescope. These data were used to constrain the orbital eccentricity to be $e < 0.039$ ($3\sigma$), but were not sufficient to determine whether WASP-24\,b possesses an atmospheric inversion layer. \citet{Smith+12aa} also observed one transit in the Str\"omgren $u$ and $y$ passbands with the BUSCA multi-band imager \citep[see][]{Me+12mn2} and provided new measurements of the physical properties of the system.

\citet{Knutson+14apj} studied the orbital motion of WASP-24 over 3.5 years using high-precision RVs from multiple telescopes. They found no evidence for orbital eccentricity or for a long-term drift attributable to a third body in the system. Finally, \citet{Sada+12pasp} obtained one transit light curve of WASP-24, and spectral analyses of the host star have been performed by \citet{Torres+12apj} and \citet{Mortier+13aa}.

\begin{figure*} \includegraphics[width=\textwidth,angle=0]{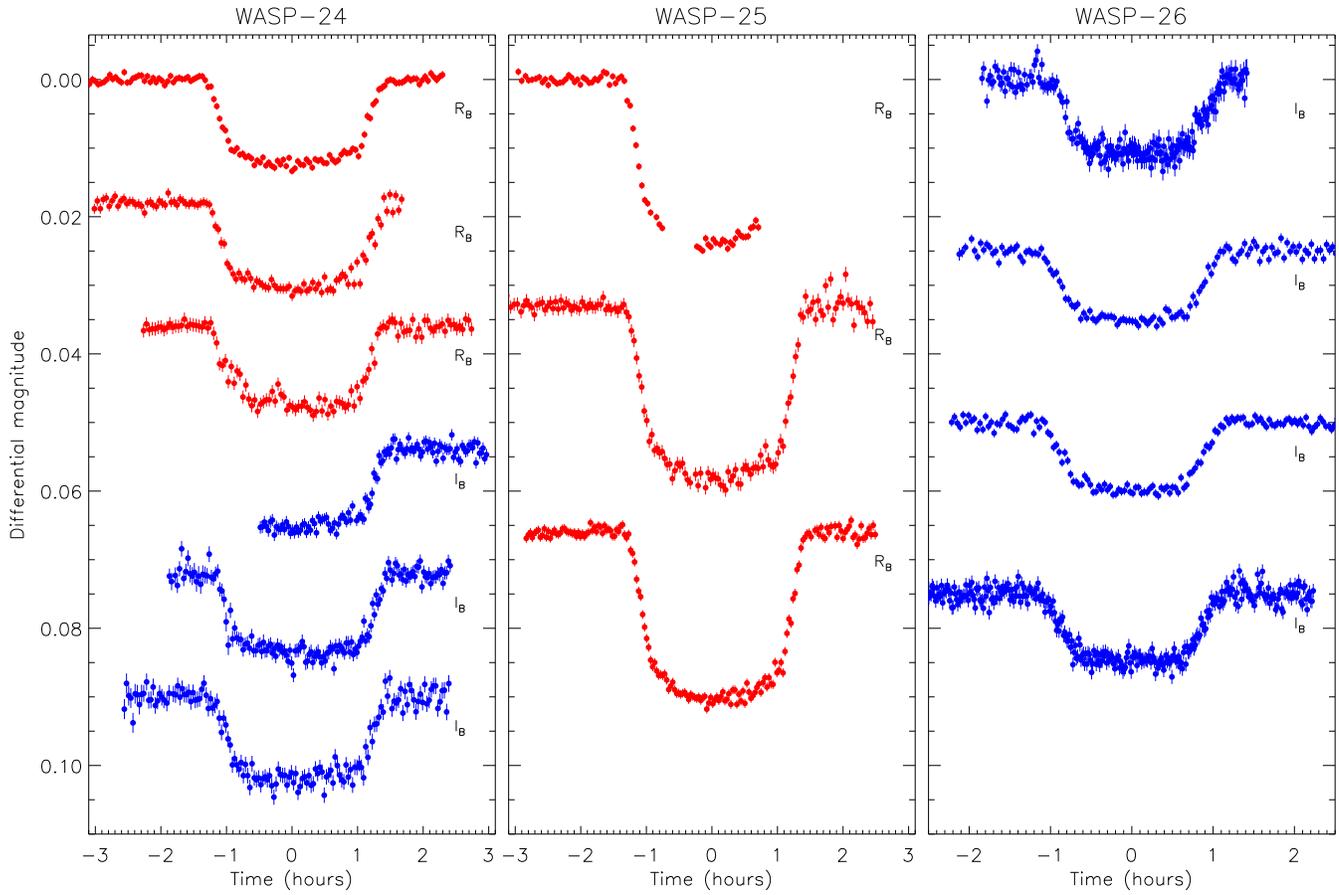}
\caption{\label{fig:lcall} Light curves presented in this work, in the
order they are given in Table\,\ref{tab:obslog}. Times are given relative
to the midpoint of each transit, and the filter used is indicated. Blue
and red filled circles represent observations through the Bessell $R$
and $I$ filters, respectively.} \end{figure*}

\subsection{WASP-25}                                                                                                        \label{sec:intro:w25}

WASP-25 \citep{Enoch+11mn} is a comparatively unstudied system containing a low-density transiting planet (mass 0.6\Mjup, radius 1.2\Rjup) orbiting a solar-like star (mass 1.1\Msun, radius 0.9\Rsun) every 3.76\,d. Their follow-up observations included two transits, one observed with FTS and one with the Euler telescope. \citet{Brown+12mn} observed one transit with HARPS, detecting the RM effect and finding $\lambda = 14.6 \pm 6.7^\circ$. They deduced that this is consistent with an aligned orbit, using the Bayesian Information Criterion.

\citet{Maxted++11mn} measured the effective temperature (\Teff) of WASP-25\,A using the infrared flux method. \citet{Mortier+13aa} obtained the spectral parameters of the star from high-resolution spectroscopy.

\subsection{WASP-26}                                                                                                        \label{sec:intro:w26}

WASP-26 was discovered by \citet{Smalley+10aa} and contains a typical hot Jupiter (mass 1.0\Mjup, radius 1.2\Rjup) orbiting a G0\,V star (mass 1.1\Msun, radius 1.3\Rsun) in a circular 2.75\,d orbit. WASP-26 has a common-proper-motion companion at 15\as\ which is roughly 2.5\,mag fainter than the planet host star. \citet{Smalley+10aa} observed one transit of WASP-26 with FTS and one with a large scatter with FTN.

\citet{Anderson+11aa} obtained high-precision RVs using HARPS through one transit of WASP-26, but their data were insufficient to allow detection of the RM effect. They also observed a transit with a 35\,cm telescope; the data are too scattered to be useful for the current work. \citet{Albrecht+12apj2} observed a spectroscopic transit using Keck/HIRES and made a low-confidence detection of the RM effect resulting in $\lambda = \er{-34}{36}{26}$\degr.

\citet{Mahtani+13mn} observed two occultations, at 3.6\,$\mu$m and 4.5\,$\mu$m, using {\it Spitzer}. They were unable to distinguish whether the planet has an atmosphere with or without a thermal inversion, but could conclude that the orbit was likely circular with $e < 0.04$ at $3\sigma$ confidence. \citet{Mahtani+13mn} also presented light curves of a transit taken in the $g$, $r$ and $i$ filters, using BUSCA.

\citet{Maxted++11mn} measured the \Teff\ of WASP-26\,A using the infrared flux method. \citet{Mortier+13aa} determined the atmospheric parameters of the star from high-resolution spectroscopy.


\section{Observations and data reduction}                                                                                             \label{sec:obs}

\subsection{Observations}

\begin{table*} \centering
\caption{\label{tab:obslog} Log of the observations presented in this work. $N_{\rm obs}$ is the number
of observations, $T_{\rm exp}$ is the exposure time, $T_{\rm dead}$ is the dead time between exposures,
`Moon illum.' is the fractional illumination of the Moon at the midpoint of the transit, and $N_{\rm poly}$
is the order of the polynomial fitted to the out-of-transit data. The aperture radii are target aperture,
inner sky and outer sky, respectively.}
\setlength{\tabcolsep}{5pt}
\begin{tabular}{lcccccccccccc} \hline
Target  & Date of   & Start time & End time  &$N_{\rm obs}$ & $T_{\rm exp}$ & $T_{\rm dead}$ & Filter & Airmass & Moon & Aperture   & $N_{\rm poly}$ & Scatter \\
        & first obs &    (UT)    &   (UT)    &              & (s)           & (s)            &        &         &illum.& radii (px) &                & (mmag)  \\
\hline
WASP-24 & 2010 06 16 & 00:29 & 06:08 & 129 & 120      & 39 & $R$ & 1.36 $\to$ 1.17 $\to$ 2.38 & 0.275 & 29 45 80 & 2 & 0.454 \\ 
WASP-24 & 2011 05 06 & 02:29 & 08:07 & 125 & 120      & 42 & $R$ & 1.47 $\to$ 1.17 $\to$ 1.77 & 0.085 & 30 40 70 & 2 & 0.745 \\ 
WASP-24 & 2011 06 29 & 00:03 & 05:03 & 113 & 120      & 40 & $R$ & 1.25 $\to$ 1.17 $\to$ 2.08 & 0.056 & 29 40 70 & 2 & 0.959 \\ 
WASP-24 & 2013 05 22 & 01:47 & 05:14 & 123 & 80--100  &  9 & $I$ & 1.37 $\to$ 1.17 $\to$ 1.26 & 0.871 & 16 28 60 & 1 & 0.825 \\ 
WASP-24 & 2013 05 29 & 00:58 & 05:15 & 151 & 80--100  & 16 & $I$ & 1.46 $\to$ 1.17 $\to$ 1.33 & 0.785 & 17 26 50 & 1 & 1.061 \\ 
WASP-24 & 2013 06 05 & 00:51 & 05:48 & 149 & 100      & 20 & $I$ & 1.37 $\to$ 1.17 $\to$ 1.63 & 0.111 & 22 30 50 & 1 & 1.185 \\ 
\hline
WASP-25 & 2010 06 13 & 23:02 & 02:42 &  72 & 100--120 & 41 & $R$ & 1.04 $\to$ 1.00 $\to$ 1.63 & 0.035 & 28 40 65 & 1 & 0.494 \\ 
WASP-25 & 2013 05 03 & 02:12 & 08:00 & 139 & 112--122 & 25 & $R$ & 1.01 $\to$ 1.00 $\to$ 2.41 & 0.417 & 20 32 70 & 2 & 1.040 \\ 
WASP-25 & 2013 06 05 & 23:58 & 05:17 & 173 & 100      &  9 & $R$ & 1.02 $\to$ 1.00 $\to$ 2.07 & 0.058 & 22 35 70 & 1 & 0.663 \\ 
\hline
WASP-26 & 2012 09 17 & 02:26 & 05:42 & 222 & 31--60   & 46 & $I$ & 1.33 $\to$ 1.03 $\to$ 1.04 & 0.017 & 20 50 80 & 1 & 1.201 \\ 
WASP-26 & 2013 08 22 & 03:39 & 08:17 & 124 & 120      & 14 & $I$ & 1.43 $\to$ 1.03 $\to$ 1.45 & 0.980 & 18 50 80 & 1 & 0.689 \\ 
WASP-26 & 2013 09 02 & 04:11 & 09:29 & 153 & 100      & 25 & $I$ & 1.17 $\to$ 1.03 $\to$ 1.47 & 0.101 & 20 55 80 & 2 & 0.621 \\ 
WASP-26 & 2013 09 12 & 03:36 & 09:16 & 393 & 30       & 25 & $I$ & 1.15 $\to$ 1.03 $\to$ 1.69 & 0.567 & 14 50 80 & 1 & 1.029 \\ 
\hline \end{tabular} \end{table*}

All observations were taken with the DFOSC (Danish Faint Object Spectrograph and Camera) instrument mounted on the 1.54\,m Danish Telescope at ESO La Silla, Chile. This setup yields a field of view of 13.7\am$\times$13.7\am\ at a plate scale of 0.39\as\,pixel$^{-1}$. We defocussed the telescope in order to improve the precision and efficiency of our observations (see \citealt{Me+09mn} for detailed signal to noise calculations). We windowed the CCD in order to lower the amount of observing time lost to readout. The autoguider was used to maintain pointing, resulting in a drift of no more than five pixels through individual observing sequences. Most nights were photometric. An observing log is given in Table\,\ref{tab:obslog} and the final light curves are plotted in Fig.\,\ref{fig:lcall}. The data were taken through either a Bessell $R$ or Bessell $I$ filter.

Two of our light curves do not have full coverage of a transit. We missed the start of the transit of WASP-24 on 2013/05/22 due to telescope pointing restrictions. Parts of the transit of WASP-25 on 2010/06/13 were lost to technical problems and then cloud. Finally, data for one transit of WASP-24 and one of WASP-26 extend only slightly beyond egress as high winds demanded closure of the telescope dome.

\subsection{Telescope and instrument upgrades}

Up to and including the 2011 observing season, the CCD in DFOSC was operated with a gain of $\sim$1.4 ADU per e$^-$, a readout noise of $\sim$4.3\,e$^-$, and 16-bit digitisation. As part of a major overhaul of the Danish telescope, a new CCD controller was installed for the 2012 season. The CCD is now operated with a much higher gain ($\sim$4.2 ADU per e$^-$) and 32-bit digitisation, so the readout noise ($\sim$5.0\,e$^-$) is much smaller relative to the number of ADU recorded for a particular star. The onset of saturation with the new CCD controller is at roughly 680\,000 ADU (M.\ I.\ Andersen, private communication).

For the current project we aimed for a maximum pixel count rate of between 250\,000 and 350\,000 ADU, in order to ensure that we stayed well below the threshold for saturation. The effect of this is that less defocussing was required due to the greater dynamic range of the CCD controller, so the object apertures for the 2012 and 2013 season are smaller than those for the 2010 and 2011 seasons. The lesser importance of readout noise also means that the CCD could be read out more quickly, so the newer data have a higher observational cadence. These effects are visible in Table\,\ref{tab:obslog}.

\subsection{Aperture photometry}

The data were reduced using the {\sc defot} pipeline, which is written in {\sc idl}\footnote{The acronym {\sc idl} stands for Interactive Data Language and is a trademark of ITT Visual Information Solutions. For further details see: {\tt http://www.ittvis.com/ProductServices/IDL.aspx}.} and uses routines from the {\sc astrolib} library\footnote{The {\sc astrolib} subroutine library is distributed by NASA. For further details see: {\tt http://idlastro.gsfc.nasa.gov/}.}. {\sc defot} has undergone several modifications since its first use \citep{Me+09mn} and we review these below.

\reff{The first modification is that pointing changes due to telescope guiding errors are measured by cross-correlating each image against a reference image, using the following procedure. Firstly, the image in question and the reference image are each collapsed in the $x$ and $y$ directions, whilst avoiding areas affected by a significant number of bad pixels. The resulting one-dimensional arrays are each divided by a robust polynomial fit, where the quantity minimised is the mean-absolute-deviation rather than the usual least-squares. The $x$ and $y$ arrays are then cross-correlated, and Gaussian functions are fit to the peaks of the cross-correlation functions in order to measure the spatial offset. The photometric apertures are then shifted by the measured amounts in order to track the motion of the stellar images across the CCD.}

This modification has been in routine use since our analysis of WASP-2 \citep{Me+10mn}. It performs extremely well as long as there is no field rotation during observations. It is much easier to track offsets between entire images rather than the alternative of following the positions of individual stars in the images, as the PSFs are highly non-Gaussian so their centroids are difficult to measure\footnote{See \citet{Nikolov+13aa} for one way of determining the centroid of a highly defocussed PSF}.

Aperture photometry was performed by the {\sc defot} pipeline using the {\sc aper} algorithm from the {\sc astrolib} implementation of the {\sc daophot} package \citep{Stetson87pasp}. We placed the apertures by hand on the target and comparison stars, and tried a wide range of sizes for all three apertures. For our final light curves we used the aperture sizes which yielded the most precise photometry, measured versus a fitted transit model (see below). We find that different choices of aperture size do affect the photometric precision but do not yield differing transit shapes. The aperture sizes are reported in Table\,\ref{tab:obslog}.

\subsection{Bias and flat-field calibrations}

Master bias and flat-field calibration frames were constructed for each observing season, by median-combining large numbers of individual bias and twilight sky images. For each observing sequence we tested whether their inclusion in the analysis produces photometry with a lower scatter. Inclusion of the master bias image was found to have a negligible effect in all cases, whereas using a master flat field can either aid or hinder the quality of the resulting photometry. It only led to a significant improvement in the scatter of the light curve for the observation of WASP-24 on 2010/06/16. It is probably not a coincidence that this dataset yielded the least scattered light curve either with or without flat-fielding.

We attribute the divergent effects of flat-fielding to the varying relative importance of the advantages and disadvantages of the calibration process. The main advantage is that variations in pixel efficiency, which occur on several spatial scales, can be compensated for. Small-scale variations (i.e.\ variations between adjacent pixels) average down to a low level as our defocussed PSFs cover of order 1000 pixels, so this effect is unimportant. Large-scale variations (e.g.\ differing illumination levels over the CCD) are usually dealt with by autoguiding the telescope -- flat-fielding is in general more important for cases when the telescope tracking is poor. The disadvantages of the standard approach to flat-fielding are:-- (1) the master flat-field image has Poisson noise which is propagated into the science images; (2) pixel efficiency depends on wavelength so observations of red stars are not properly calibrated using observations of a blue twilight sky; (3) pixel efficiency depends on the number of counts, which is in general different for the science and the calibration observations.

\subsection{Light curve generation}

The instrumental magnitudes of the target and comparison stars were converted into differential-magnitude light curves normalised to zero magnitude outside transit, using the following procedure. For each observing sequence an ensemble comparion star was constructed by adding the fluxes of all good comparison stars with weights adjusted to give the lowest possible scatter for the data taken outside transit. The normalisation was performed by fitting a polynomial to the out-of-transit datapoints. We used a first-order polynomial when possible, as this cannot modify the shape of the transit, but switched to a second-order polynomial when the observations demanded. The weights of the comparison stars and the coefficients were optimised simultaneously to yield the final differential-magnitude light curve. The order of the polynomial used for each dataset is given in Table\,\ref{tab:obslog}.

In the original version of the {\sc defot} pipeline the optimisation of the weights and coefficients was performed using the {\sc idl} {\sc amoeba} routine, which is an implementation of the downhill simplex algorithm of \citet{NelderMead65}. We have found that this routine can suffer from irreproducibility of results, primarily as it is prone to getting trapped in local minima. We have therefore modified {\sc defot} to use the {\sc mpfit} implementation of the Levenberg-Marquardt algorithm \citep{Markwardt07aspc}. We find the fitting process to be much faster and more reliable when using {\sc mpfit} compared to using {\sc amoeba}.

The timestamps for the datapoints have been converted to the BJD(TDB) timescale \citep{Eastman++10pasp}. Manual time checks were obtained for several frames and the FITS file timestamps were confirmed to be on the UTC system to within a few seconds. The timings therefore appear not to suffer from the same problems as previously found for WASP-18 and \reff{suspected for} WASP-16 \citep{Me+09apj,Me+13mn}. The light curves are shown in Fig.\,\ref{fig:lcall}. The reduced data are ennumerated in Table\,\ref{tab:lcdata} and will be made available at the CDS\footnote{{\tt http://vizier.u-strasbg.fr/}}.

\begin{table} \centering \caption{\label{tab:lcdata} Excerpts of the light curves
presented in this work. The full dataset will be made available at the CDS.}
\begin{tabular}{lccrr} \hline
Target & Filter & BJD(TDB) & Diff.\ mag. & Uncertainty \\
\hline
WASP-24 & $R$ & 2455364.525808 & -0.00079 & 0.00054 \\
WASP-24 & $R$ & 2455364.527590 & -0.00018 & 0.00053 \\
WASP-24 & $R$ & 2455364.529430 & -0.00001 & 0.00055 \\[2pt]
WASP-25 & $R$ & 2455361.464729 & -0.00113 & 0.00052 \\
WASP-25 & $R$ & 2455361.466882 &  0.00019 & 0.00051 \\
WASP-25 & $R$ & 2455361.469289 & -0.00049 & 0.00051 \\[2pt]
WASP-26 & $I$ & 2456187.608599 & -0.00016 & 0.00101 \\
WASP-26 & $I$ & 2456187.609444 & -0.00161 & 0.00103 \\
WASP-26 & $I$ & 2456187.611111 &  0.00316 & 0.00103 \\
\hline \end{tabular} \end{table}


\section{High-resolution imaging}                                                                                                      \label{sec:li}

\begin{figure*}
\includegraphics[width=0.325\textwidth,angle=0]{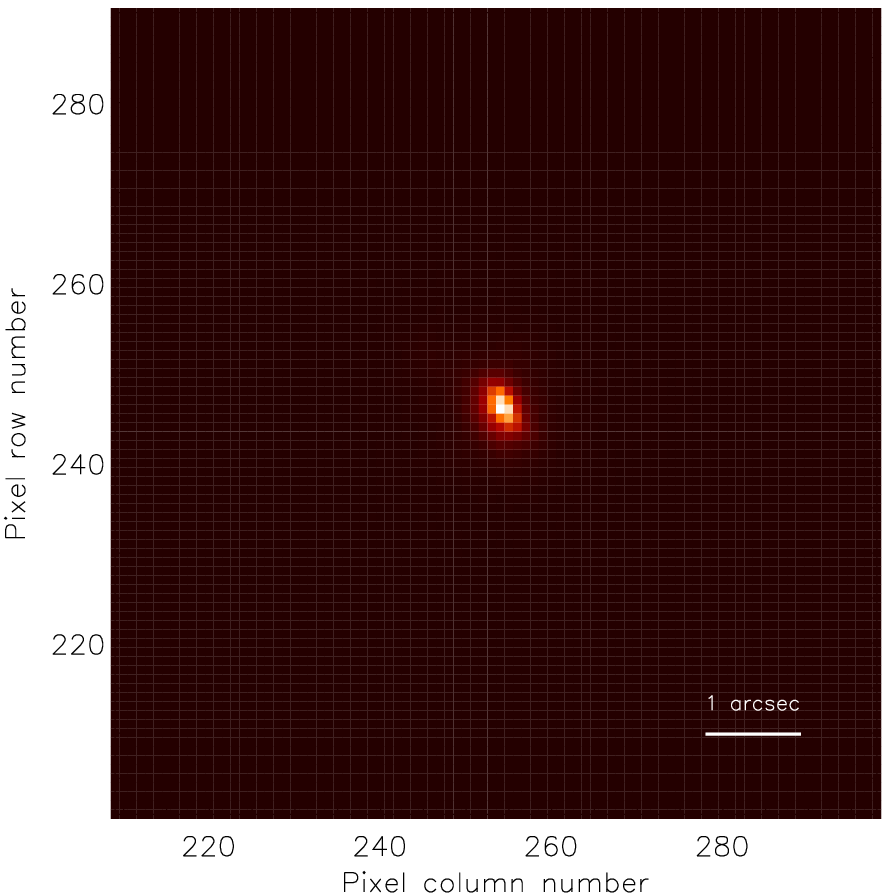}
\includegraphics[width=0.325\textwidth,angle=0]{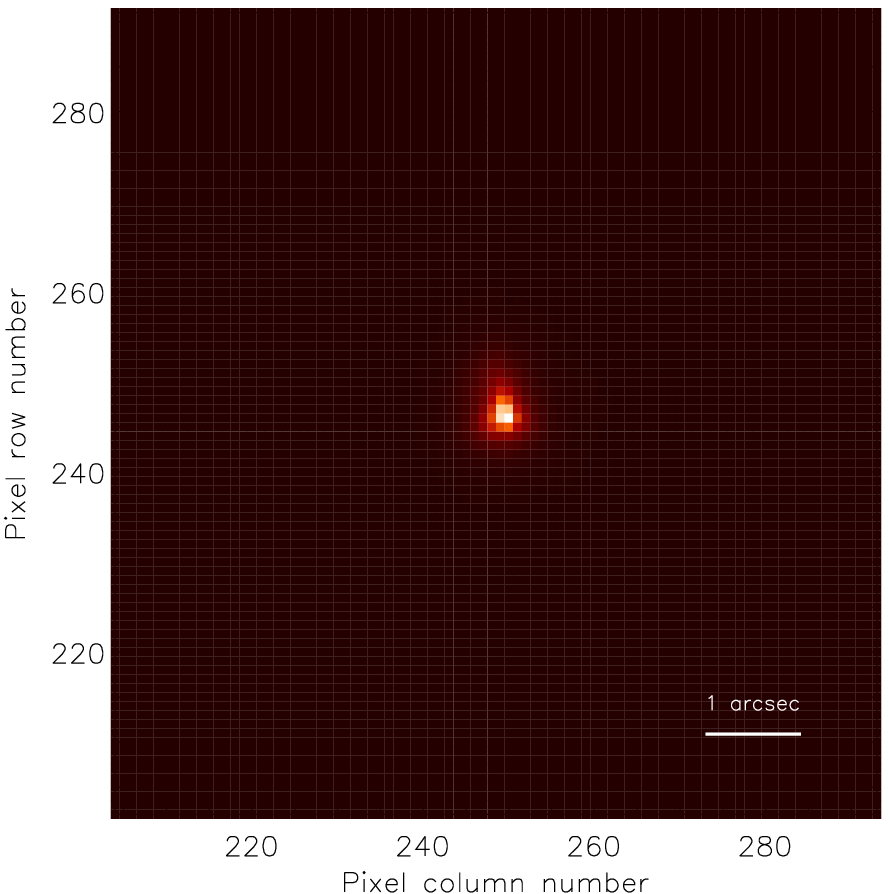}
\includegraphics[width=0.325\textwidth,angle=0]{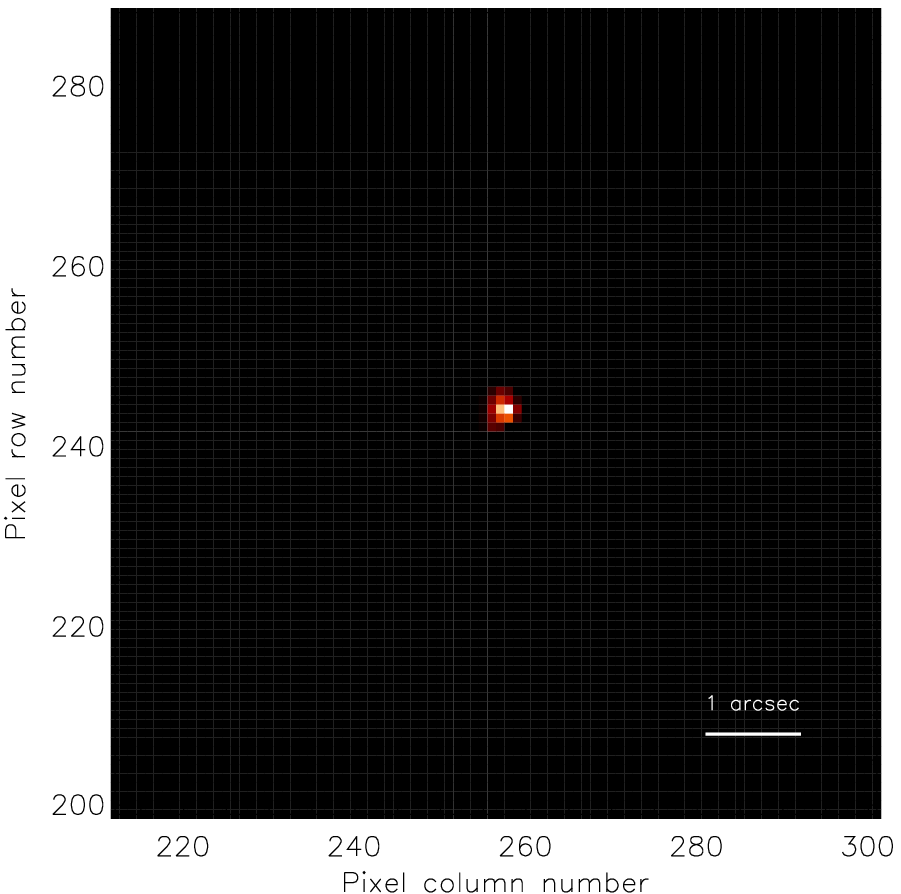}
\caption{\label{fig:li:lin} \reff{High-resolution Lucky Imaging observations
of WASP-24 (left), WASP-25 (middle) and WASP-26 (right). In each case an image
covering $8\as \times 8\as$ and centred on our target star is shown. A bar of
length $1\as$ is superimposed in the bottom-right of each image. The flux scale
is linear. Each image is a sum of the best 2\% of the original images, so the
effective exposure times are 2.4\,s, 4.4\,s and 2.1\,s, respectively.}} \end{figure*}

\begin{figure*}
\includegraphics[width=0.325\textwidth,angle=0]{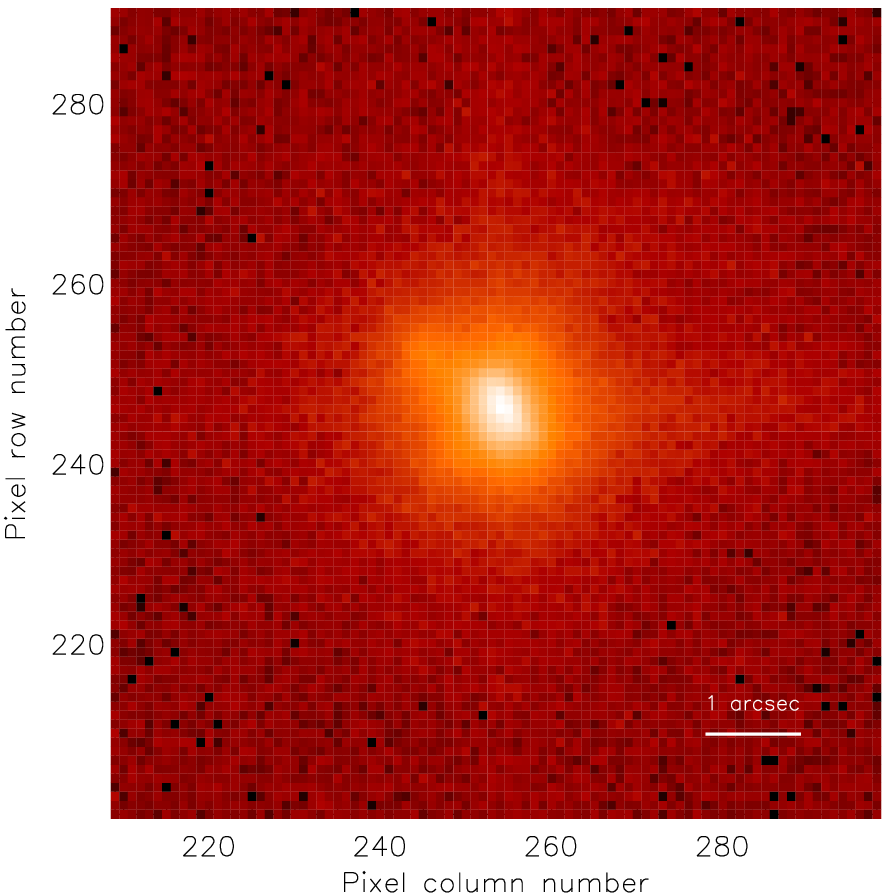}
\includegraphics[width=0.325\textwidth,angle=0]{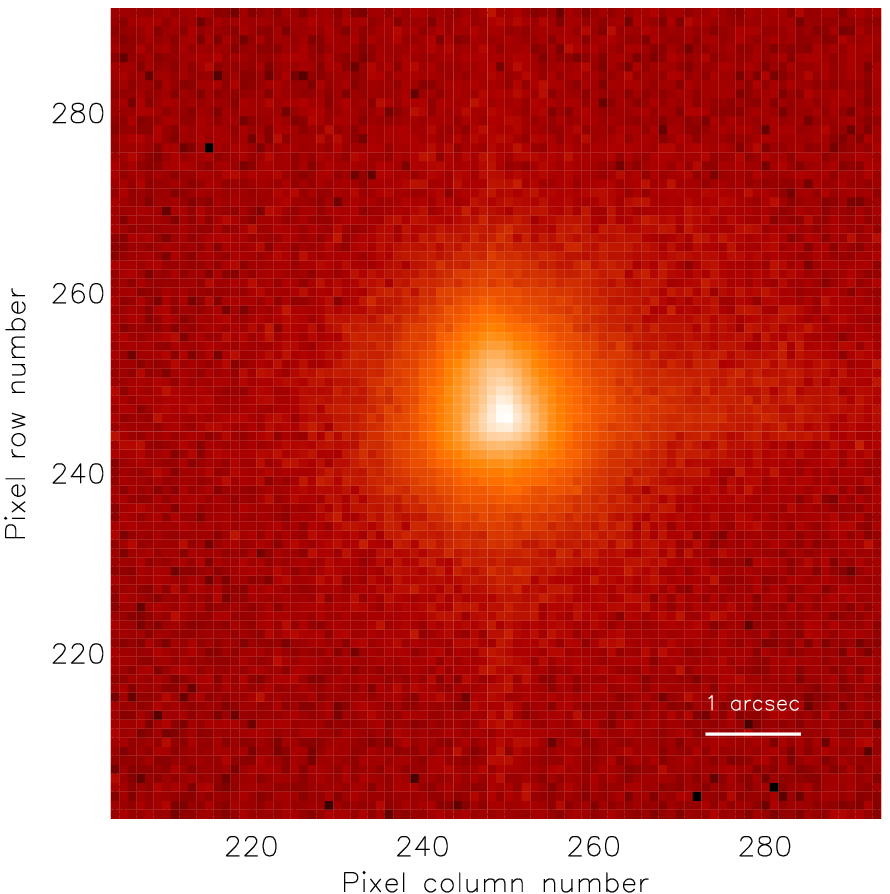}
\includegraphics[width=0.325\textwidth,angle=0]{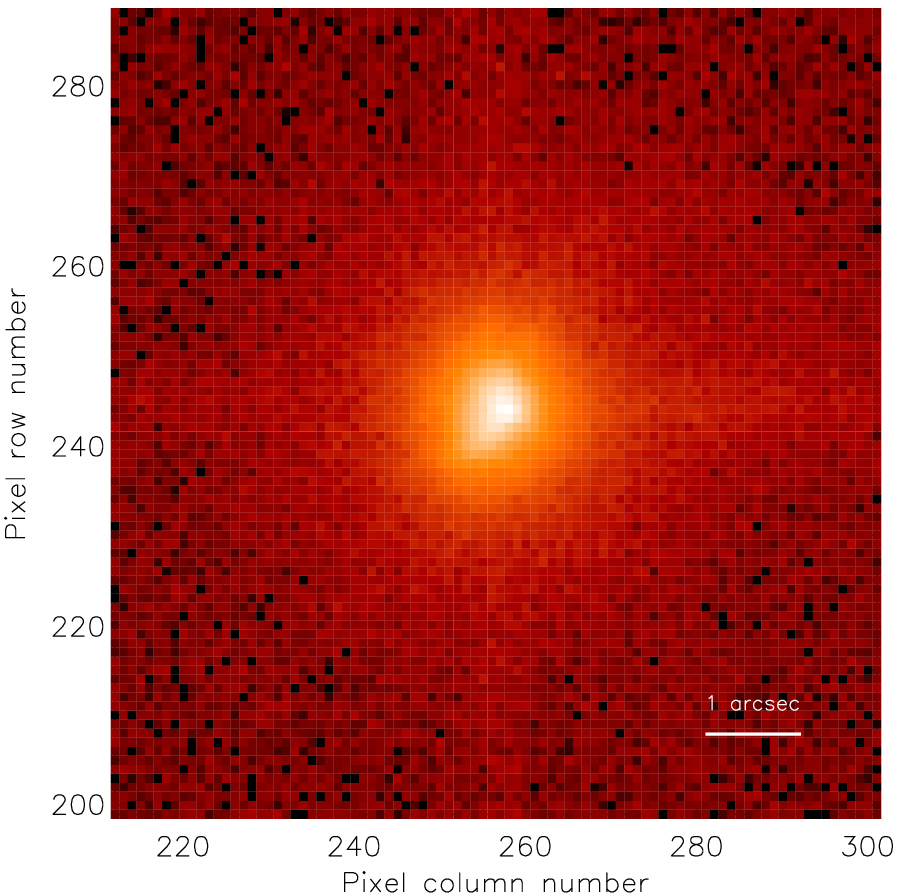}
\caption{\label{fig:li:log} \reff{Same as Fig.\,\ref{fig:li:lin}, except that the flux
scale is logarithmic so faint stars are more easily identified.}} \end{figure*}



For each object we obtained well-focussed images with DFOSC in order to check for faint nearby stars whose light might have contaminated that from our target star. Such objects would dilute the transit and cause us to underestimate the radius of the planet \citep{Daemgen+09aa}. The worst-case scenario is a contaminant which is an eclipsing binary, as this would render the planetary nature of the system questionable.

\reff{For WASP-24 we find nearby stars at 43 and 55 pixels (16.8\as\ and 21.5\as), which are more than 7.6 and 4.5 mag fainter than the target star in the $R$ filter. Precise photometry is not available for the focussed images as WASP-24 itself is saturated to varying degrees. We estimate that the star at 43 pixels contributes less than 0.01\% of the flux in the inner aperture of WASP-24, which is much too small to affect our results. The star at 55 pixels is an eclipsing binary (see Section\,\ref{sec:eb}) but its PSF was always clearly separated from that of WASP-24 so it also contributes an unmeasurably small amount of flux to the inner aperture of WASP-24.}

\reff{For WASP-25 the nearest star is at 94 pixels and is 5.36\,mag fainter than our target. The inner aperture for WASP-25 is significantly smaller than this distance, so the presence of the nearby star has a negligible effect on our photometry.} For WASP-26 there is a known star which is 39 pixels (15.2\as) away from the target and 2.55\,mag fainter in our images. The object and sky apertures in Section\,\ref{sec:obs} were selected such that this star was in no-man's land between them, and thus had an insignificant effect on our photometry.

In order to search for stars which are very close to our target systems, we obtained high-resolution images of \reff{all three targets} using the Lucky Imager (LI) mounted on the Danish telescope. The LI uses an Andor 512$\times$512 pixel electron-multiplying CCD, with a pixel scale of 0.09\as\,pixel$^{-1}$ and a field of view of $45\as\times45\as$. The data were reduced using a dedicated pipeline and the best 2\% of images were stacked together to yield combined images whose PSF is smaller than the seeing limit. A long-pass filter was used, resulting in a response which approximates that of SDSS $i$$+$$z$ \citep{Skottfelt+13aa}. \reff{Exposure times of 120\,s, 220\,s and 109\,s were used for WASP-24, WASP-25 and WASP-26, respectively.} The LI observations are thus shallower than the focussed DFOSC images, but have a better resolution. A detailed examination of different high-resolution imaging approaches was recently given by \citet{Lillobox++14aa}

The central parts of the images are shown in Figs.\ \ref{fig:li:lin} and \ref{fig:li:log}. The image for WASP-24 has a PSF FWHM of 4.1\,px in $x$ (pixel column) and 4.8\,px in $y$ (pixel row), corresponding to 0.37\as$\times$0.43\as. The image of WASP-25 is nearly as good (4.2$\times$5.3 pixels), \reff{and that for WASP-26 is better (3.8$\times$4.4 pixels).} None of the images show any stars which were undetected on our focussed DFOSC observations, so we find no evidence for contaminating light in the PSFs of the targets. There is a suggestion of a very faint star north-east of WASP-24, but this was not confirmed by a repeat image. If present, its brightness is insufficient to have a significant effect on our analysis.



\section{Orbital period determination}                                                                                               \label{sec:porb}

\begin{table*} \begin{center}
\caption{\label{tab:minima} Times of minimum light and their residuals versus the ephemeris derived in this work.}
\begin{tabular}{l l l r r l} \hline
Target & Time of minimum & Uncertainty & Cycle & Residual & Reference \\
       & (BJD/TDB)) & (d) & number & (d) &       \\
\hline
WASP-24 & 2455081.38018 & 0.00017 &  -259.0 &  0.00044 & \citet{Street+10apj}         \\   
WASP-24 & 2455308.47842 & 0.00151 &  -162.0 &  0.00017 & Ayiomamitis (TRESCA)         \\   
WASP-24 & 2455308.48020 & 0.00163 &  -162.0 &  0.00195 & Br\'at (TRESCA)              \\   
WASP-24 & 2455322.52496 & 0.00074 &  -156.0 & -0.00061 & This work (BUSCA $u$-band)   \\   
WASP-24 & 2455322.52498 & 0.00049 &  -156.0 & -0.00059 & This work (BUSCA $y$-band)   \\   
WASP-24 & 2455364.66718 & 0.00024 &  -138.0 & -0.00038 & This work (Danish Telescope) \\   
WASP-24 & 2455687.75622 & 0.00038 &     0.0 &  0.00006 & This work (Danish Telescope) \\   
WASP-24 & 2455701.80338 & 0.00049 &     6.0 & -0.00011 & \citet{Sada+12pasp}          \\   
WASP-24 & 2455741.60468 & 0.00052 &    23.0 &  0.00042 & This work (Danish Telescope) \\   
WASP-24 & 2456010.84351 & 0.00052 &   138.0 & -0.00125 & Wallace et al.\ (TRESCA)     \\   
WASP-24 & 2456010.84412 & 0.00062 &   138.0 & -0.00064 & Wallace et al.\ (TRESCA)     \\   
WASP-24 & 2456408.85005 & 0.00257 &   308.0 & -0.00240 & Garlitz (TRESCA)             \\   
WASP-24 & 2456441.63058 & 0.00042 &   322.0 &  0.00102 & This work (Danish Telescope) \\   
WASP-24 & 2456448.65324 & 0.00049 &   325.0 &  0.00002 & This work (Danish Telescope) \\   
\hline
WASP-25 & 2455274.99726 & 0.00021 &  -163.0 &  0.00015 & \citet{Enoch+11mn}       \\   
WASP-25 & 2455338.99804 & 0.00075 &  -146.0 & -0.00123 & Curtis (TRESCA)          \\   
WASP-25 & 2455659.01066 & 0.00118 &   -61.0 &  0.00061 & Curtis (TRESCA)          \\   
WASP-25 & 2455677.83276 & 0.00078 &   -56.0 & -0.00145 & Evans (TRESCA)           \\   
WASP-25 & 2456415.74114 & 0.00021 &   140.0 & -0.00028 & This work                \\   
WASP-25 & 2456430.80140 & 0.00063 &   144.0 &  0.00065 & Evans (TRESCA)           \\   
WASP-25 & 2456449.62499 & 0.00012 &   149.0 &  0.00008 & This work                \\   
\hline
WASP-26 & 2455123.63867 & 0.00070 &  -259.0 & -0.00086 & \citet{Smalley+10aa}     \\   
WASP-26 & 2455493.02404 & 0.00183 &  -125.0 &  0.00048 & Curtis (TRESCA)          \\   
WASP-26 & 2456187.68731 & 0.00043 &   127.0 &  0.00125 & This work                \\   
WASP-26 & 2456526.74716 & 0.00041 &   250.0 & -0.00036 & This work                \\   
WASP-26 & 2456537.77389 & 0.00036 &   254.0 & -0.00002 & This work                \\   
WASP-26 & 2456548.79992 & 0.00038 &   258.0 & -0.00038 & This work                \\   
\hline \end{tabular} \end{center} \end{table*}

\begin{figure*}
\includegraphics[width=\textwidth,angle=0]{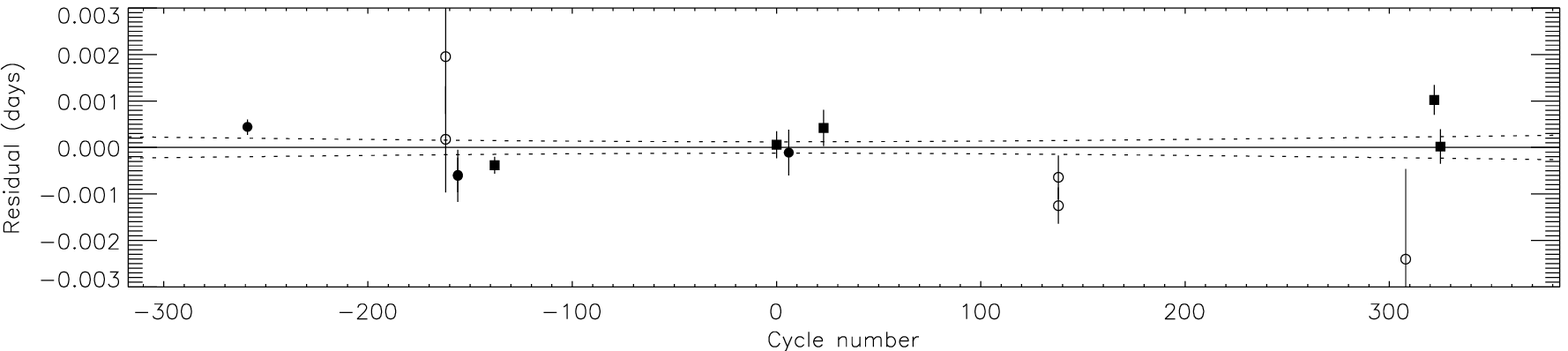}
\includegraphics[width=\textwidth,angle=0]{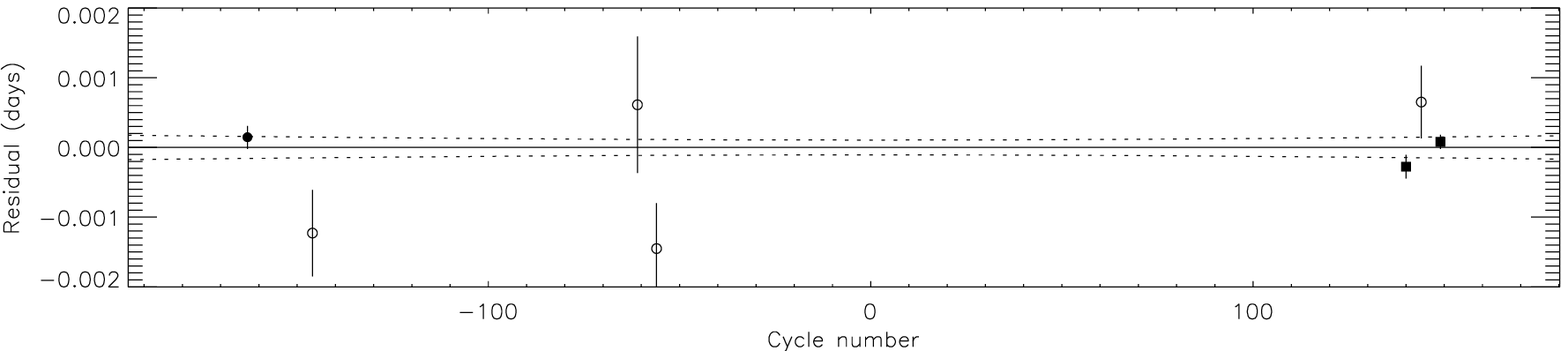}
\includegraphics[width=\textwidth,angle=0]{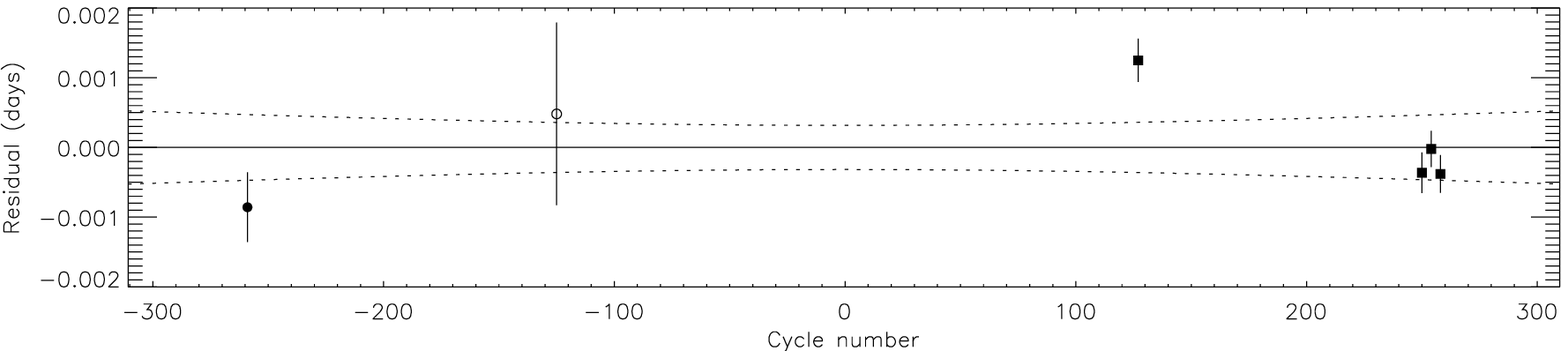}
\caption{\label{fig:minima} Plot of the residuals of the timings of mid-transit
versus a linear ephemeris, for WASP-24 (top), WASP-25 (middle) and WASP-26 (bottom).
The results from this work are shown using filled squares, and from amateur observers
with open circles. All other timings are shown by filled circles. The dotted lines show
the 1$\sigma$ uncertainty in the ephemeris as a function of cycle number. \reff{The
errorbars have been scaled up to force $\chi^2_\nu = 1.0$.}} \end{figure*}

Our first step was to improve the measured orbital ephemerides of the three TEPs using our new data. Each of our light curves was fitted using the {\sc jktebop} code (see below) and their errorbars were rescaled to give a reduced $\chi^2$ of $\chi^2_\nu = 1.0$ versus the fitted model. This step is necessary as the uncertainties from the {\sc aper} algorithm tend to be underestimated. We then fitted each revised dataset to measure the transit midpoints and ran Monte Carlo simulations to estimate the uncertainties in the midpoints. The two transits with only partial coverage were not included in this analysis, as they yield less reliable timings \citep[e.g.][]{Gibson+09apj}.

We have collected additional times of transit midpoint from literature sources. Those from the discovery papers \citep{Street+10apj,Enoch+11mn,Smalley+10aa} are on the UTC timescale (D.\ R.\ Anderson, private communication) so we converted them to TDB to match our own results. We used the timings from our own fits to the BUSCA light curves presented by \citet{Smith+12aa} for WASP-24.

We also collated minimum timings from the Exoplanet Transit Database\footnote{The Exoplanet Transit Database (ETD) can be found at: {\tt http://var2.astro.cz/ETD/credit.php}} \citep{Poddany++10newa}, which provides data and times of minimum from amateur observers affiliated with TRESCA\footnote{The TRansiting ExoplanetS and CAndidates (TRESCA) website can be found at: {\tt http://var2.astro.cz/EN/tresca/index.php}}. We retained only those timing measurements based on light curves where all four contact points of the transit are easily identifiable by eye. We assumed that the times were all on the UTC timescale and converted them to TDB.

For each object we fitted the times of mid-transit with straight lines to determine new linear orbital ephemerides. Table\,\ref{tab:minima} gives all transit times plus their residual versus the fitted ephemeris. The uncertainties have been increased to force $\chi^2_\nu = 1.0$, $E$ gives the cycle count versus the reference epoch, and the bracketed numbers show the uncertainty in the final digit of the preceding number.

The revised ephemeris for WASP-24 is:
$$ T_0 = {\rm BJD(TDB)} \,\, 2\,455\,687.75616 (16) \, + \, 2.3412217 (8) \times E $$
where the errorbars have been inflated to account for $\chi^2_\nu = 1.75$. We have adopted one of our timings from the 2011 season as the reference epoch. This is close to the midpoint of the available data so the covariance between the orbital period and the time of reference epoch is small.

Our orbital ephemeris for WASP-25 is:
$$ T_0 = {\rm BJD(TDB)} \,\, 2\,455\,888.66484 (13) \, + \, 3.7648327 (9) \times E $$
accounting for $\chi^2_\nu = 1.21$. We have adopted a reference epoch midway between our 2013 data and the timing from the discovery paper.

The new orbital ephemeris for WASP-26 is:
$$ T_0 = {\rm BJD(TDB)} \,\, 2\,455\,837.59821 (44) \, + \, 2.7565972 (19) \times E $$
accounting for $\chi^2_\nu = 1.40$ and using a reference epoch in mid-2011. The main contributor to the  $\chi^2_\nu$ is our transit from 2012, which was observed under conditions of poor sky transparency. Whilst a parabolic ephemeris provides a formally better fit to the transit times, this improvement is due almost entirely to our 2012 transit so is not reliable.

Fig.\,\ref{fig:minima} shows the residuals versus the linear ephemeris for each of our three targets. No transit timing variations are discernable by eye, and there are insufficient timing measurements to perform a quantitative search for such variations. Our period values for all three systems are consistent with previous measurements but are significantly more precise due to the longer temporal baseline of the available transit timings.


\section{Light curve analysis}                                                                                                         \label{sec:lc}

We have analysed the light curves using the {\it Homogeneous Studies} methodology (see \citealt{Me12mn} and references therein), which utilises the {\sc jktebop}\footnote{{\sc jktebop} is written in {\sc fortran77} and the source code is available at {\tt http://www.astro.keele.ac.uk/jkt/codes/jktebop.html}} code \citep{Me++04mn} and the NDE model \citep{NelsonDavis72apj,PopperEtzel81aj}. This represents the star and planet as spheres for the calculation of eclipse shapes and as biaxial spheroids for proximity effects.

The fitted parameters of the model for each system were the fractional radii of the star and planet ($r_{\rm A}$ and $r_{\rm b}$), the orbital inclination ($i$), limb darkening coefficients, and the reference time of mid-transit. The fractional radii are the ratio between the true radii and the semimajor axis: $r_{\rm A,b}= \frac{R_{\rm A,b}}{a}$. They were expressed as their sum and ratio, $r_{\rm A} + r_{\rm b}$ and $k = \frac{r_{\rm b}}{r_{\rm A}}$, because these two quantities are more weakly correlated. The orbital period was held fixed at the value found in Section\,\ref{sec:porb}. We assumed a circular orbit for each system based on the case histories given in Section\,\ref{sec:intro}.

Whilst the light curves had already been rectified to zero differential magnitude outside transit, the uncertainties in this process need to be propagated through subsequent analyses. This effect is relatively unimportant for transits with plenty of data before ingress and after egress, as the rectification polynomial is well-defined and needs only to be interpolated to the data within transit. It is, however, crucial for partial transits as the rectification polynomial is defined on only a short stretch of data on one side of the transit, which then needs to be extrapolated to all in-transit data. {\sc jktebop} was therefore modified to allow multiple polynomials to be specified, each operating on only a subset of data within a specific time interval. This allowed multiple light curves to be modelled simultaneously but subject to independent polynomial fits to the out-of-transit data. For each transit we included as fitted parameters the coefficients of a polynomial of order given in Table\,\ref{tab:obslog}. \reff{We found that the coefficients of the polynomials did not exhibit strong correlations against the other model parameters: the correlation coefficients are normally less than 0.4.}

Limb darkening (LD) was accounted for by each of five LD laws \citep[see][]{Me08mn}, with the linear coefficients either fixed at theoretically predicted values\footnote{Theoretical LD coefficients were obtained by bilinear interpolation to the host star's \Teff\ and \logg\ using the {\sc jktld} code available from: {\tt http://www.astro.keele.ac.uk/jkt/codes/jktld.html}} or included as fitted parameters. We did not calculate fits for both LD coefficients in the four bi-parametric laws as they are very strongly correlated \citep{Me08mn,Carter+08apj}. The nonlinear coefficients were instead perturbed by $\pm$0.1 on a flat distribution during the error analysis simulations, in order to account for imperfections in the theoretically predicted coefficients.

Error estimates for the fitted parameters were obtained in several ways. We ran solutions using different LD laws, and also calculated errorbars using residual-permutation and Monte Carlo algorithms \citep{Me08mn}. The final value for each parameter is the unweighted mean of the four values from the solutions using the two-parameter LD laws. Its errorbar was taken to be the larger of the Monte-Carlo or residual-permutation alternatives, with an extra contribution to account for variations between solutions with the different LD laws. Tables of results for each light curve, including our reanalysis of published data, can be found in the Supplementary Information.

\subsection{Results for WASP-24}

\begin{table*} \caption{\label{tab:w24:lcfit} Parameters of the
fit to the light curves of WASP-24 from the {\sc jktebop} analysis (top). The
final parameters are given in bold and the parameters found by other studies
are shown (below). Quantities without quoted uncertainties were not given
by those authors but have been calculated from other parameters which were.}
\begin{tabular}{l r@{\,$\pm$\,}l r@{\,$\pm$\,}l r@{\,$\pm$\,}l r@{\,$\pm$\,}l r@{\,$\pm$\,}l}
\hline
Source       & \mc{$r_{\rm A}+r_{\rm b}$} & \mc{$k$} & \mc{$i$ ($^\circ$)} & \mc{$r_{\rm A}$} & \mc{$r_{\rm b}$} \\
\hline
Danish Telescope $R$-band & 0.1900 & 0.0057 & 0.1029 & 0.0013 & 83.60 & 0.50 & 0.1723 & 0.0050 & 0.01773 & 0.00070 \\
Danish Telescope $I$-band & 0.1805 & 0.0077 & 0.1012 & 0.0013 & 84.23 & 0.72 & 0.1639 & 0.0068 & 0.01659 & 0.00086 \\
Street LT/RISE            & 0.2028 & 0.0157 & 0.1080 & 0.0044 & 82.85 & 1.31 & 0.1830 & 0.0135 & 0.01976 & 0.00213 \\
Street FTN                & 0.1807 & 0.0138 & 0.1011 & 0.0014 & 84.12 & 1.21 & 0.1642 & 0.0123 & 0.01659 & 0.00138 \\
Sada \reff{KPNO} $J$-band & 0.2146 & 0.0370 & 0.1090 & 0.0063 & 81.34 & 2.53 & 0.1935 & 0.0327 & 0.02108 & 0.00433 \\
Smith BUSCA $u$-band      & 0.1556 & 0.1203 & 0.0990 & 0.0061 & 86.57 & 3.43 & 0.1416 & 0.0212 & 0.01401 & 0.00312 \\
Smith BUSCA $y$-band      & 0.1766 & 0.0173 & 0.1016 & 0.0032 & 84.59 & 1.90 & 0.1603 & 0.0156 & 0.01630 & 0.00189 \\
\hline
Final results&{\bf0.1855}&{\bf0.0042}&{\bf0.1018}&{\bf0.0007}&{\bf83.87}&{\bf0.38}&{\bf0.1684}&{\bf0.0037}&{\bf0.01713}&{\bf0.00049}\\
\hline
\citet{Street+10apj}      & \mc{0.1866} & 0.1004 & 0.0006 & 83.64 & 0.31 & \mc{0.1696} & \mc{0.01702} \\
\citet{Smith+12aa}        & \mc{0.1922} & 0.1050 & 0.0006 & 83.30 & 0.30 & 0.1739 & 0.0033 & \mc{0.01826} \\
\hline \end{tabular} \end{table*}

\begin{figure} \includegraphics[width=\columnwidth,angle=0]{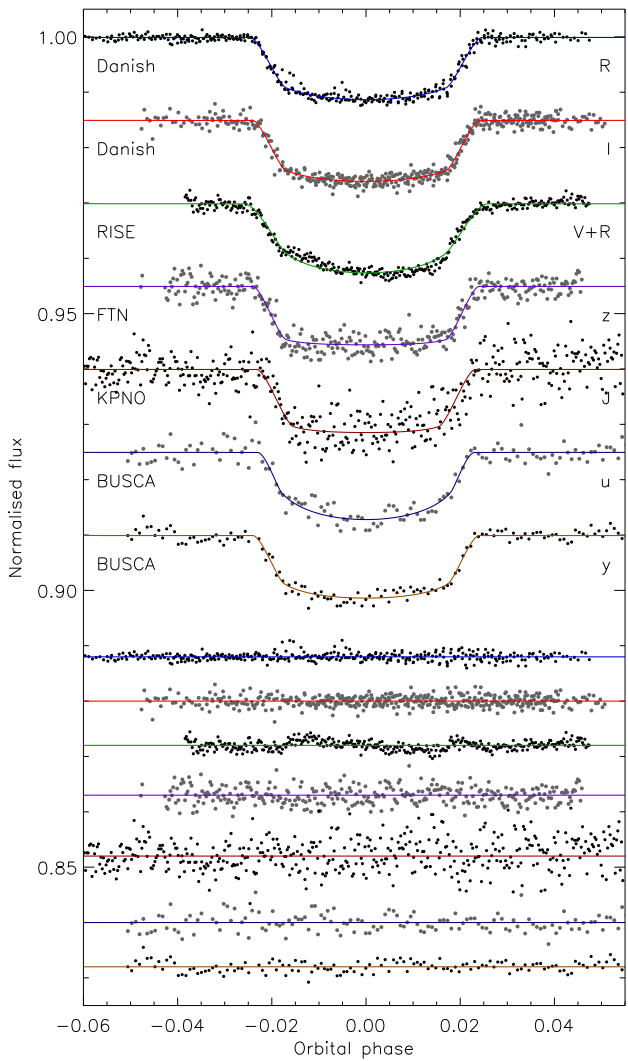}
\caption{\label{fig:w24:lc} The phased light curves of WASP-24 analysed in this
work, compared to the {\sc jktebop} best fits. The residuals of the fits
are plotted at the base of the figure, offset from unity. Labels give the
source and passband for each dataset. The polynomial baseline functions
have been removed from the data before plotting.} \end{figure}

For WASP-24 we divided our data into two datasets, one for the $R$ and one for the $I$ filters. For each we calculated solutions for all five LD laws under two scenarios: both LD coefficients fixed (`LD-fixed'), and the linear coefficient fitted whilst the nonlinear coefficient was fixed but then perturbed in the error analysis simulations (`LD-fit/fix'). The two datasets give consistent results and show no signs of red noise (the Monte Carlo errorbars were similar to or larger than the residual-permutation errorbars).

We also modelled published transit light curves of WASP-24. The discovery paper \citep{Street+10apj} presented two light curves which covered complete transits, one from the RISE instrument on the LT and one using Merope on the FTN. The RISE data were first binned by a factor of 10 from 3454 to 346 datapoints to lower the required CPU time. \citet{Sada+12pasp} observed one transit in the $J$ band with the KPNO 2.1\,m telescope. \citet{Smith+12aa} obtained photometry of one transit simultaneously in the Str\"omgren $u$ and $y$ bands.

We found that red noise was strong in the RISE and KPNO data (see Fig.\,\ref{fig:w24:lc}) so the results from these datasets were not included in our final values. The $u$-band data gave exceptionally uncertain results so we also discounted this dataset. The photometric results from the LD-fit/fix cases for the remaining four datasets \reff{were} combined according to weighted means, to obtain the final photometric parameters of WASP-24 (Table\,\ref{tab:w24:lcfit}). We also checked what the values would be had we not rejected any combination of the three least reliable datasets, and found changes of less than half the errorbars in all cases.

Table\,\ref{tab:w24:lcfit} also shows a comparison between our values and literature results. We note that the two previous publications gave inconsistent results (see in particular the respective values for $k$) despite being based on much of the same data. This implies that their error estimates were optimistic. To obtain final values for the photometric parameters of WASP-24 we have calculated the weighted mean of those from individual datasets. The results found in the current work are based on more extensive data and analysis, and should be preferred over previous values.

\subsection{Results for WASP-25}

\begin{table*} \caption{\label{tab:w25:lcfit} Parameters of the
fit to the light curves of WASP-25 from the {\sc jktebop} analysis (top). The
final parameters are given in bold and the parameters found by other studies
are shown (below). Quantities without quoted uncertainties were not given
by \citet{Enoch+11mn} but have been calculated from other parameters which were.}
\begin{tabular}{l r@{\,$\pm$\,}l r@{\,$\pm$\,}l r@{\,$\pm$\,}l r@{\,$\pm$\,}l r@{\,$\pm$\,}l}
\hline
Source       & \mc{$r_{\rm A}+r_{\rm b}$} & \mc{$k$} & \mc{$i$ ($^\circ$)} & \mc{$r_{\rm A}$} & \mc{$r_{\rm b}$} \\
\hline
Danish Telescope  & 0.1004 & 0.0019 & 0.1384 & 0.0011 & 88.33 & 0.32 & 0.0882 & 0.0016 & 0.01221 & 0.00030 \\
Enoch FTS         & 0.1072 & 0.0043 & 0.1416 & 0.0026 & 87.54 & 0.52 & 0.0939 & 0.0036 & 0.01328 & 0.00067 \\
Enoch Euler       & 0.1004 & 0.0067 & 0.1374 & 0.0029 & 88.13 & 1.37 & 0.0883 & 0.0056 & 0.01214 & 0.00098 \\
\hline
Final results&{\bf0.1015}&{\bf0.0017}&{\bf0.1387}&{\bf0.0010}&{\bf88.12}&{\bf0.27}&{\bf0.0891}&{\bf0.0014}&{\bf0.01237}&{\bf0.00028}\\
\hline
\citet{Enoch+11mn} & \mc{0.1029} & 0.1367 & 0.0007 & 88.0 & 0.5 & \mc{0.09049} & \mc{0.01237} \\
\hline \end{tabular} \end{table*}

\begin{figure} \includegraphics[width=\columnwidth,angle=0]{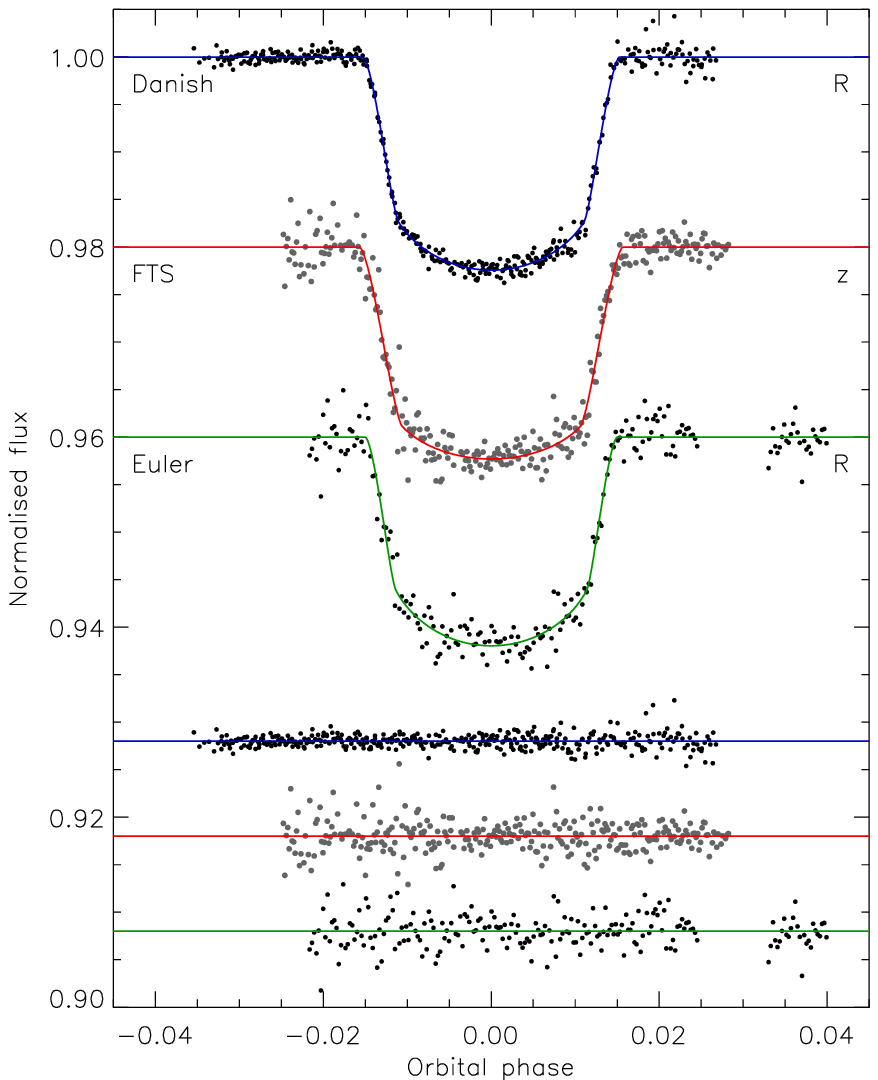}
\caption{\label{fig:w25:lc} The phased light curves of WASP-25 analysed in this
work, compared to the {\sc jktebop} best fits. The residuals of the fits
are plotted at the base of the figure, offset from unity. Labels give the
source and passband for each dataset. The plynomial baseline functions
have been removed from the data before plotting.} \end{figure}

Our three transits were all taken in the Bessell $R$ band so were modelled together. We found that red noise was not important and that the data contained sufficient information to fit for the linear LD coefficient. The best fits are plotted in Fig.\,\ref{fig:w25:lc}

\citet{Enoch+11mn} obtained two transit light curves of WASP-25 in their initial characterisation of this object, one from FTS with the Spectral camera and one from the Swiss Euler telescope with EulerCam. For both datasets we have adopted the LD-fit/fix values. The FTS data have significant curvature outside transit, implying that a quadratic baseline should be included. If this is done then $r_{\rm A}+r_{\rm b}$ and $k$ become smaller by approximately 1$\sigma$ and $i$ greater by 1.5$\sigma$, yielding the values in Table\,\ref{tab:w25:lcfit}. This change is significantly larger than the errorbars quoted by \citet{Enoch+11mn}, which are based primarily on the FTS and the less precise Euler data.

\subsection{Results for WASP-26}

\begin{table*} \caption{\label{tab:w26:lcfit} Parameters of the
fit to the light curves of WASP-26 from the {\sc jktebop} analysis (top). The
final parameters are given in bold and the parameters found by other studies
are shown (below). Quantities without quoted uncertainties were not given
by those authors but have been calculated from other parameters which were.}
\begin{tabular}{l r@{\,$\pm$\,}l r@{\,$\pm$\,}l r@{\,$\pm$\,}l r@{\,$\pm$\,}l r@{\,$\pm$\,}l}
\hline
Source       & \mc{$r_{\rm A}+r_{\rm b}$} & \mc{$k$} & \mc{$i$ ($^\circ$)} & \mc{$r_{\rm A}$} & \mc{$r_{\rm b}$} \\
\hline
Danish Telescope & 0.1584 & 0.0044 & 0.0973 & 0.0008 & 83.29 & 0.32 & 0.1444 & 0.0040 & 0.01405 & 0.00038 \\
Smalley FTS      & 0.176  & 0.011  & 0.1027 & 0.0044 & 82.47 & 0.63 & 0.160  & 0.010  & 0.0164 & 0.0016 \\
Mahtani $g$-band & 0.1733 & 0.0089 & 0.1081 & 0.0029 & 82.31 & 0.53 & 0.1564 & 0.0077 & 0.0169 & 0.0012 \\
Mahtani $r$-band & 0.174  & 0.017  & 0.1026 & 0.0042 & 82.6  & 1.2  & 0.158  & 0.015  & 0.0162 & 0.0016 \\
Mahtani $i$-band & 0.184  & 0.035  & 0.103  & 0.032  & 81.5  & 2.2  & 0.166  & 0.020  & 0.0172 & 0.0091 \\
\hline
Final results&{\bf0.1649}&{\bf0.0040}&{\bf0.0991}&{\bf0.0018}&{\bf82.83}&{\bf0.27}&{\bf0.1505}&{\bf0.0036}&{\bf0.01465}&{\bf0.00054}\\
\hline
\citet{Smalley+10aa}  & \mc{0.1716} & 0.101  & 0.002  & 82.5 & 0.5 & \mc{0.1559} & \mc{0.01574} \\
\citet{Anderson+11aa} & \mc{0.1675} & 0.1011 & 0.0017 & 82.5 & 0.5 & \mc{0.1521} & \mc{0.01538} \\
\citet{Mahtani+13mn}  & \mc{0.1661} & 0.1015 & 0.0015 & 82.5 & 0.5 & \mc{0.1508} & \mc{0.01536} \\
\hline \end{tabular} \end{table*}

\begin{figure} \includegraphics[width=\columnwidth,angle=0]{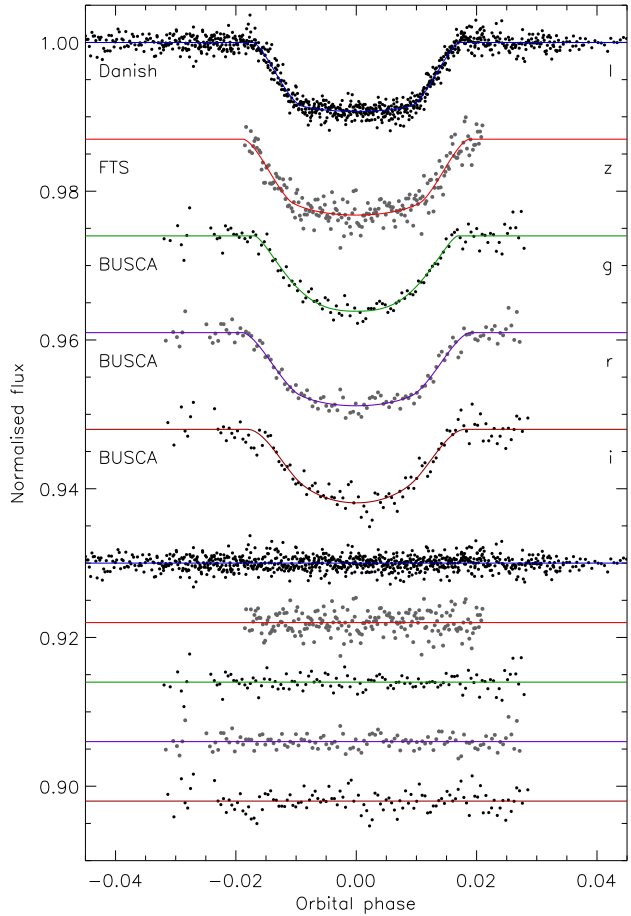}
\caption{\label{fig:w26:lc} The phased light curves of WASP-26 analysed in this
work, compared to the {\sc jktebop} best fits. The residuals of the fits
are plotted at the base of the figure, offset from unity. Labels give the
source and passband for each dataset. The polynomial baseline functions
have been removed from the data before plotting.} \end{figure}

The four transits presented in this work were all taken in the Bessell $I$ band, so were modelled together. We found once again that red noise was not important and that the data contained sufficient information to fit for the linear LD coefficient. The best fit is shown in Fig.\,\ref{fig:w26:lc} and the parameter values are given in Table\,\ref{tab:w26:lcfit}.

\citet{Smalley+10aa} obtained two transit light curves, one each from FTS/Spectral and FTN/Merope. The former has almost no out-of-transit data, and the latter is very scattered. We modelled the FTS light curve here but did not attempt to extract information from the FTN data. We found that the scatter was dominated by white noise and it was not possible to fit for any LD coefficients.

\citet{Mahtani+13mn} presented photometry of one transit of WASP-26 obtained simultaneously in the $g$, $r$ and $i$ bands using BUSCA. We modelled these datasets individually. The $g$- and $r$-band data could only support a LD-fixed solution. Red noise was unimportant for $g$ and $r$ but the residual-permutation errorbars were a factor of 2.5 greater than the Monte Carlo errorbars for $i$.

Table\,\ref{tab:w26:lcfit} collects the parameter values found from each light curve. The data from the Danish Telescope are of much higher precision than previous datasets, and yield a solution with larger orbital inclination and smaller fractional radii than obtained in previous studies. Whilst $r_{\rm A}+r_{\rm b}$ and $i$ are in overall agreement ($\chir = 1.0$ and $0.8$ versus the weighted mean value), $k$ and $r_{\rm B}$ are not ($\chir = 3.4$ and $1.8$). These moderate discrepancies were accounted for by increasing the errorbars on the final weighted-mean parameter values, \reff{by an amount sufficient to force $\chir = 1.0$}.


\section{Physical properties}


\begin{table} \centering \caption{\label{tab:spec} Spectroscopic properties of the
planet host stars used in the determination of the physical properties of the systems.
\newline {\bf References:}
(1) \citet{Torres+12apj};
(2) \citet{Knutson+14apj};
(3) \citet{Mortier+13aa};
(4) \citet{Enoch+10aa};
(5) \citet{Maxted++11mn};
(6) \citet{Smalley+10aa};
(7) \citet{Mahtani+13mn}}
\begin{tabular}{l r@{\,$\pm$\,}l r@{\,$\pm$\,}l r@{\,$\pm$\,}l c}
\hline
Target & \mc{\Teff\ (K)} & \mc{\FeH\ (dex)} & \mc{$K_{\rm A}$ (\ms)} & Ref \\
\hline
WASP-24 & 6107 & 77 & $-$0.02 & 0.10 & 152.1 & 3.2 & 1,1,2 \\
WASP-25 & 5736 & 50 &    0.06 & 0.05 &  75.5 & 5.3 & 3,3,4 \\
WASP-26 & 6015 & 55 & $-$0.02 & 0.09 & 138   &   2 & 5,6,7 \\
\hline \end{tabular} \end{table}

We have measured the physical properties of the three planetary systems using the photometric quantities \reff{found} in Section \ref{sec:lc}, published spectroscopic results, and five sets of theoretical stellar evolutionary models \citep{Claret04aa,Demarque+04apjs,Pietrinferni+04apj,Vandenberg++06apjs,Dotter+08apjs}. Table\,\ref{tab:spec} gives the spectroscopic quantities adopted from the literature, where $K_{\rm A}$ denotes the velocity amplitude of the star.

In the case of WASP-24 there are two recent conflicting spectroscopic analyses: \citet{Torres+12apj} measured $\Teff = 6107 \pm 77$\,K and $\logg = 4.26 \pm 0.01$ (c.g.s.) whereas \citet{Mortier+13aa} obtained $\Teff = 6297 \pm 58$\,K and $\logg = 4.76 \pm 0.17$. We have adopted the former \Teff\ as it agrees with an independent value from \citet{Street+10apj} and the corresponding \logg\ is in good agreement with that derived from our own analysis.

For each object we used the measured values of $r_{\rm A}$, $r_{\rm b}$, $i$ and $K_{\rm A}$, and an estimated value of the velocity amplitude of the planet, $K_{\rm b}$, to calculate the physical properties of the system. $K_{\rm b}$ was then iteratively refined to obtain the best agreement between the calculated $\frac{R_{\rm A}}{a}$ and the measured $r_{\rm A}$, and between the spectroscopic \Teff\ and that predicted by the stellar models for the observed \FeH\ and the calculated stellar mass ($M_{\rm A}$). This was done for a range of ages in order to determine the overall best fit and age of the system. Further details on the method can be found in \citet{Me09mn}. This process was performed for each of the five sets of theoretical stellar models, in order to estimate the systematic error incurred by the use of stellar theory.

\begin{table*} \caption{\label{tab:model} Derived physical properties of the three systems.
\reff{Where two sets of errorbars are given, the first is the statistical uncertainty and the second is the systematic uncertainty.}}
\begin{tabular}{l l l r@{\,$\pm$\,}c@{\,$\pm$\,}l r@{\,$\pm$\,}c@{\,$\pm$\,}l r@{\,$\pm$\,}c@{\,$\pm$\,}l} \hline
Quantity                & Symbol           & Unit    & \mcc{WASP-24}               & \mcc{WASP-25}               & \mcc{WASP-26}               \\
\hline
Stellar mass            & $M_{\rm A}$      & \Msun  & 1.168   & 0.056   & 0.050   & 1.053   & 0.023   & 0.030   & 1.095   & 0.043   & 0.017   \\
Stellar radius          & $R_{\rm A}$      & \Rsun  & 1.317   & 0.036   & 0.019   & 0.924   & 0.016   & 0.009   & 1.284   & 0.035   & 0.007   \\
Stellar surface gravity & $\log g_{\rm A}$ & c.g.s. & 4.267   & 0.021   & 0.006   & 4.530   & 0.014   & 0.004   & 4.260   & 0.022   & 0.002   \\
Stellar density         & $\rho_{\rm A}$   & \psun  & \mcc{$0.512 \pm 0.034$}     & \mcc{$1.336 \pm 0.063$}     & \mcc{$0.517 \pm 0.037$}     \\[2pt]
Planet mass             & $M_{\rm b}$      & \Mjup  & 1.109   & 0.043   & 0.032   & 0.598   & 0.044   & 0.012   & 1.020   & 0.031   & 0.011   \\
Planet radius           & $R_{\rm b}$      & \Rjup  & 1.303   & 0.043   & 0.019   & 1.247   & 0.030   & 0.012   & 1.216   & 0.047   & 0.006   \\
Planet surface gravity  & $g_{\rm b}$      & \mss   & \mcc{$16.19 \pm  0.99$}     & \mcc{$9.54 \pm 0.80$}       & \mcc{$17.1 \pm  1.3$}       \\
Planet density          & $\rho_{\rm b}$   & \pjup  & 0.469   & 0.042   & 0.007   & 0.288   & 0.028   & 0.003   & 0.530   & 0.060   & 0.003   \\[2pt]
Equilibrium temperature & \Teq\            & K      & \mcc{$1772 \pm   29$}       & \mcc{$1210 \pm   14$}       & \mcc{$1650 \pm   24$}       \\
Safronov number         & \safronov\       &        & 0.0529  & 0.0021  & 0.0007  & 0.0439  & 0.0033  & 0.0004  & 0.0607  & 0.0026  & 0.0003  \\
Orbital semimajor axis  & $a$              & au     & 0.03635 & 0.00059 & 0.00052 & 0.04819 & 0.00035 & 0.00046 & 0.03966 & 0.00052 & 0.00021 \\
Age                     & $\tau$  & (Gyr) &\ermcc{2.5}{9.6}{1.5}{1.8}{2.5} & \ermcc{0.1}{5.7}{0.1}{0.2}{0.0} & \ermcc{4.0}{5.7}{4.5}{1.4}{4.0} \\
\hline \end{tabular} \end{table*}

The final physical properties of the three planetary systems are given in Table\,\ref{tab:model}. 
The equilibrium temperatures of the planets were calculated ignoring the effects of albedo and heat redistribution:
$\Teq = \Teff \sqrt{\frac{r_{\rm A}}{2}}$.
For each parameter which depends on theoretical models there are five different values, one from using each of the five model sets. In these cases we give two errorbars: the statistical uncertainty (calculated by propagating the random errors via a perturbation analysis) and the systematic uncertainty (the maximum deviation between the final value and the five values from using the different stellar models).

The intermediate results for each set of stellar models are given in Tables A16, A17 and A18, along with a comparison to published values. We find that literature values are in generally good agreement with our own, despite being based on much less extensive follow-up photometry (see Figs.\ \ref{fig:w24:lc}, \ref{fig:w25:lc} and \ref{fig:w26:lc}) and less precise spectroscopic properties for the host stars. The uncertainties in the radii of WASP-25\,b and WASP-26\,b are significantly improved by our new results. The uncertainties in the host star mass and semimajor axis measurements for WASP-24\,b and WASP-25\,b have a significant contribution from the differences in the theoretical model predictions we used, an issue which was not considered in previous studies of these objects.


\section{Eclipsing binary star systems near WASP-24 and WASP-26}                                                                       \label{sec:eb}

\begin{figure} \includegraphics[width=\columnwidth,angle=0]{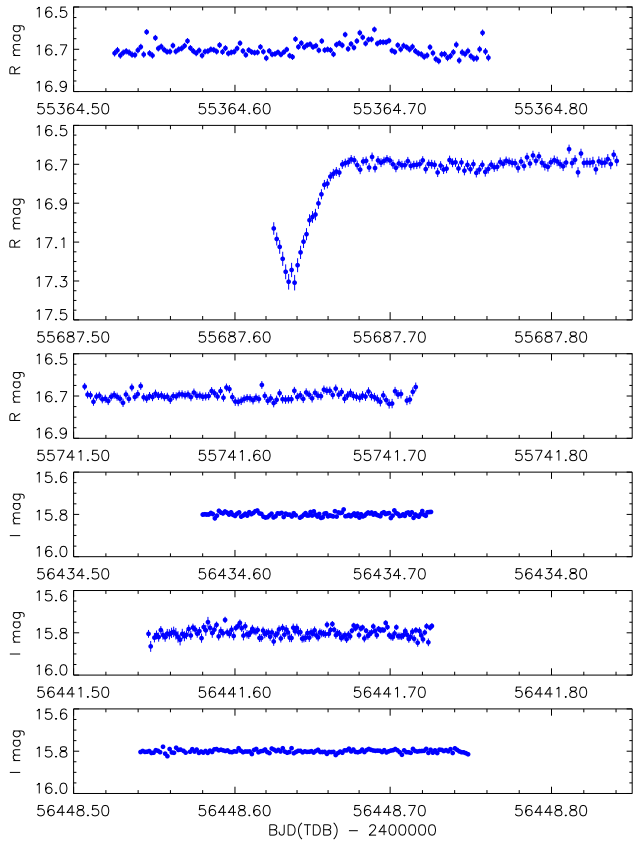}
\caption{\label{fig:w24:eb} The light curves of the eclipsing binary system near
WASP-24 from our observations. Each light curve has been shifted to an out-of-eclipse
magnitude of $R = 16.7$ \citep{Zacharias+05aas} or $I = 15.8$.} \end{figure}


\citet{Street+10apj} found the closest detected star to WASP-24 (21.2\as) to be a detached eclipsing binary system. It showed eclipses of depth 0.8\,mag in four of their follow-up photometric datasets, suggesting an orbital period of 1.156\,d. Its faintness ($V = 17.97$) means it was not measurable in the SuperWASP images. We observed one eclipse, on the night of 2011/05/05 (Fig.\,\ref{fig:w24:eb}). This confirms the eclipsing nature of the object, but is not helpful in deducing its orbital period. Further observations of this eclipsing binary would be useful in pinning down the mass-radius relation for low-mass main sequence stars \citep[e.g.][]{Lopez07apj,Torres++10aarv}.

\begin{figure} \includegraphics[width=\columnwidth,angle=0]{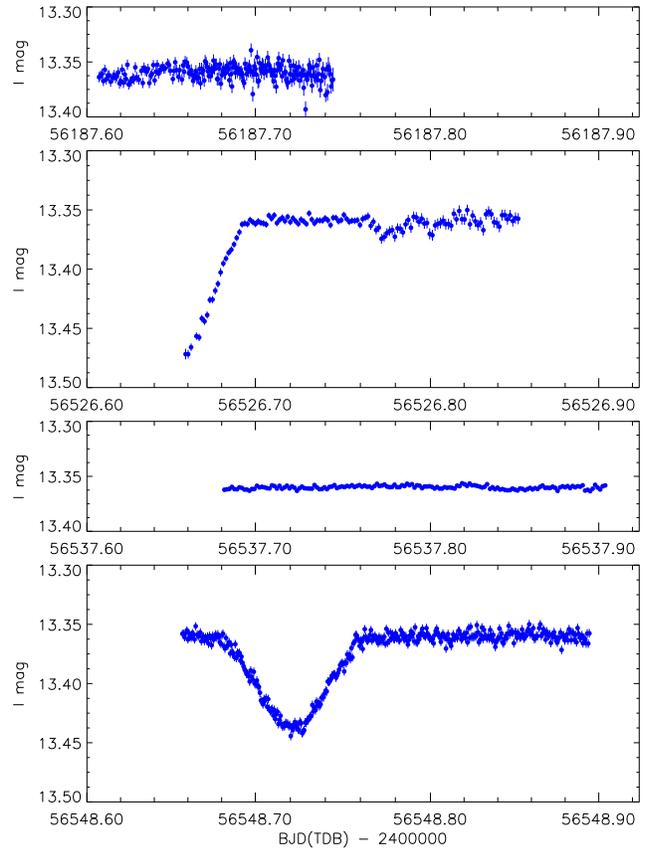}
\caption{\label{fig:w26:eb} The light curves of the eclipsing binary system
near WASP-26 from our observations. Each light curve has been shifted to an
out-of-eclipse magnitude of $I = 13.36$, calculated from its spectral type
and observed $V$ magnitude.} \end{figure}

%

In two of our datasets for WASP-26 we detected eclipses on one object which appears to be a previously unknown detached eclipsing binary system. Its sky position is approximately RA $=$ 00:18:26.5, Dec $=$ $-$15:11:49 (J2000). The AAVSO Photometric All-Sky Survey gives apparent magnitudes of $B = 16.02 \pm 0.07$ and $V = 14.98 \pm 0.02$ \citep{Henden+12javso}. The Two Micron All-Sky Survey lists it under the designation 2MASS\,J00182645$-$1511492 \citep{Skrutskie+06aj}, and its colour of $J-K = 0.72$ implies a spectral type of approximately K4\,V \citep{Currie+10apjs}. The object is not listed in the General Catalogue of Variable Stars (GCVS\footnote{{\tt http://www.sai.msu.su/gcvs/gcvs/}}) or the AAVSO Variable Star Index (VSX\footnote{{\tt http://www.aavso.org/vsx/}}).

Two eclipses were seen in the 2MASS\,J00182645$-$1511492 system, separated by approximately 22.1\,days. The first was only partially observed and has a depth of at least 0.11\,mag, whereas the full duration of the second eclipse was seen, with a depth of 0.08\,mag. The different depths mean that the former is a primary and the latter a secondary eclipse. The orbital period cannot be determined from these data, but is likely quite short as the eclipses do not last long. The SuperWASP survey \citep{Pollacco+06pasp} has obtained 5800 observations of this object, but these show no obvious variability due to the faintness of the object and the shallowness of the eclipses. Whilst it would be a useful probe of the properties of stars on the lower main sequence, 2MASS\,J00182645$-$1511492 is not a particularly promising object for further study due to its shallow eclipses, which makes the measurement of precise photometric parameters difficult, and unknown orbital period.


\section{Summary and conclusions}                                                                                                 \label{sec:summary}

\begin{figure} \includegraphics[width=0.48\textwidth,angle=0]{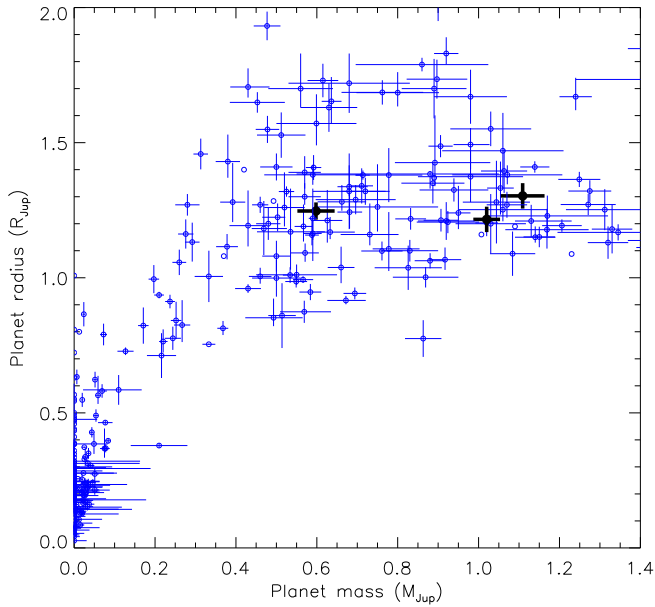}
\caption{\label{fig:m2r2} Plot of planet radii versus their masses.
WASP-24\,b, WASP-25\,b and WASP-26\,b are indicated using black filled
circles. The overall population of planets is shown using blue open
circles, using data taken from TEPCat on 2014/02/15. Errorbars are
suppressed for clarity if they are larger than 0.2\Mjup\ or 0.2\Rjup.
The outlier with a mass of 0.86\Mjup\ but a radius of only 0.78\Rjup\
is the recently-discovered system WASP-59 \citep{Hebrard+13aa}.} \end{figure}

We have presented extensive photometric observations of three Southern hemisphere transiting planetary systems discovered by SuperWASP. All three systems have spectroscopic measurements of the RM effect which are consistent with orbital alignment; two have also been observed with {\it Spitzer}. Our observations of the third, WASP-25, comprise the first follow-up photometry of this object since its discovery paper.

Our data cover thirteen transits of the gas giant planets in front of their host stars, plus single-epoch high-resolution images taken with a Lucky Imaging camera. From these observations, and published spectroscopic measurements, we have measured the orbital ephemerides and physical properties of the systems to high precision. Care was taken to propagate random errors for all quantities and assess separate statistical errors for those quantities whose evaluation depends on the use of theoretical stellar models. Previously published studies of all three objects are in good agreement with our refined values, although we find evidence that their error estimates are unrealistically small.

We have observed one eclipse for the known eclipsing binary very close to WASP-24, and discovered a new K4\,V detached eclipsing binary 4.25\,arcmin north of WASP-26. We have observed part of one primary eclipse and a full secondary eclipse for the latter object, but are not able to measure its orbital period from these observations.

Fig.\,\ref{fig:m2r2} shows a plot of planet radius versus mass for all known TEPs (data taken from the TEPCat\footnote{The Transiting Extrasolar Planet Catalogue (TEPCat) is available at: {\tt http://www.astro.keele.ac.uk/jkt/tepcat/}} catalogue on 2014/02/15). WASP-24\,b and WASP-26\,b are representative of the dominant population of Hot Jupiters, with masses near 1.0\Mjup. WASP-25\,b appears near the midpoint of a second cluster of planets with masses of approximately 0.5--0.7\Mjup; such objects are sometimes termed ``Hot Saturns'' although they are more massive than Saturn itself (0.3\Mjup).

All three planets have radii greater than predicted by theoretical models for gaseous bodies without a heavy-element core \citep{Bodenheimer++03apj,Fortney++07apj,Baraffe++08aa} so exhibit the inflated radii commonly observed for Hot Jupiters \citep[e.g.][and references therein]{Enoch+12aa}. Its deep transit and low surface gravity make WASP-25\,b a good candidate for transmission photometry and spectroscopy to probe the atmospheric properties of a transiting gas giant planet \citep[see][]{Bento+14mn}.


\section*{Acknowledgements}

%
The operation of the Danish 1.54m telescope is financed by a grant to UGJ from the Danish Natural Science Research Council.
The reduced light curves presented in this work will be made available at the CDS ({\tt http://vizier.u-strasbg.fr/}) and at {\tt http://www.astro.keele.ac.uk/$\sim$jkt/}.
J\,Southworth acknowledges financial support from STFC in the form of an Advanced Fellowship.
The research leading to these results has received funding from the European Community's Seventh Framework Programme (FP7/2007-2013/) under grant agreement Nos.\ 229517 and 268421. Funding for the Stellar Astrophysics Centre (SAC) is provided by The Danish National Research Foundation.
This publication was supported by grants NPRP 09-476-1-078 and NPRP X-019-1-006 from Qatar National Research Fund (a member of Qatar Foundation).
TCH acknowledges financial support from the Korea Research Council for Fundamental Science and Technology (KRCF) through the Young Research Scientist Fellowship Program and is supported by the KASI (Korea Astronomy and Space Science Institute) grant 2012-1-410-02/2013-9-400-00.
SG, XW and XF acknowledge the support from NSFC under the grant No.\,10873031.
The research is supported by the ASTERISK project (ASTERoseismic Investigations with SONG and Kepler) funded by the European Research Council (grant agreement No.\,267864).
DR, YD, AE, FF (ARC), OW (FNRS research fellow) and J\,Surdej acknowledge support from the Communaut\'e fran\c{c}aise de Belgique - Actions de recherche concert\'ees - Acad\'emie Wallonie-Europe.
The following internet-based resources were used in research for this paper: the ESO Digitized Sky Survey; the NASA Astrophysics Data System; the SIMBAD database and VizieR catalogue access tool operated at CDS, Strasbourg, France; and the ar$\chi$iv scientific paper preprint service operated by Cornell University.
We thank the anonymous referee for a helpful report.

\bibliographystyle{mn_new}

\begin{thebibliography}{57}
\expandafter\ifx\csname natexlab\endcsname\relax\def\natexlab#1{#1}\fi

\bibitem[{{Albrecht} et~al.(2012)}]{Albrecht+12apj2}
{Albrecht}, S., et~al., 2012, ApJ, 757, 18

\bibitem[{{Anderson} et~al.(2011)}]{Anderson+11aa}
{Anderson}, D.~R., et~al., 2011, A\&A, 534, A16

\bibitem[{{Baraffe} et~al.(2008){Baraffe}, {Chabrier}, \&
  {Barman}}]{Baraffe++08aa}
{Baraffe}, I., {Chabrier}, G., {Barman}, T., 2008, A\&A, 482, 315

\bibitem[{{Bento} et~al.(2014)}]{Bento+14mn}
{Bento}, J., et~al., 2014, MNRAS, 437, 1511

\bibitem[{{Bodenheimer} et~al.(2003){Bodenheimer}, {Laughlin}, \&
  {Lin}}]{Bodenheimer++03apj}
{Bodenheimer}, P., {Laughlin}, G., {Lin}, D.~N.~C., 2003, ApJ, 592, 555

\bibitem[{{Brown} et~al.(2012)}]{Brown+12mn}
{Brown}, D.~J.~A., et~al., 2012, MNRAS, 423, 1503

\bibitem[{{Carter} et~al.(2008){Carter}, {Yee}, {Eastman}, {Gaudi}, \&
  {Winn}}]{Carter+08apj}
{Carter}, J.~A., {Yee}, J.~C., {Eastman}, J., {Gaudi}, B.~S., {Winn}, J.~N.,
  2008, ApJ, 689, 499

\bibitem[{{Claret}(2004)}]{Claret04aa}
{Claret}, A., 2004, A\&A, 424, 919

\bibitem[{{Currie} et~al.(2010)}]{Currie+10apjs}
{Currie}, T., et~al., 2010, ApJS, 186, 191

\bibitem[{{Daemgen} et~al.(2009){Daemgen}, {Hormuth}, {Brandner}, {Bergfors},
  {Janson}, {Hippler}, \& {Henning}}]{Daemgen+09aa}
{Daemgen}, S., {Hormuth}, F., {Brandner}, W., {Bergfors}, C., {Janson}, M.,
  {Hippler}, S., {Henning}, T., 2009, A\&A, 498, 567

\bibitem[{{Demarque} et~al.(2004){Demarque}, {Woo}, {Kim}, \&
  {Yi}}]{Demarque+04apjs}
{Demarque}, P., {Woo}, J.-H., {Kim}, Y.-C., {Yi}, S.~K., 2004, ApJS, 155, 667

\bibitem[{{Dotter} et~al.(2008){Dotter}, {Chaboyer}, {Jevremovi{\'c}},
  {Kostov}, {Baron}, \& {Ferguson}}]{Dotter+08apjs}
{Dotter}, A., {Chaboyer}, B., {Jevremovi{\'c}}, D., {Kostov}, V., {Baron}, E.,
  {Ferguson}, J.~W., 2008, ApJS, 178, 89

\bibitem[{{Eastman} et~al.(2010){Eastman}, {Siverd}, \&
  {Gaudi}}]{Eastman++10pasp}
{Eastman}, J., {Siverd}, R., {Gaudi}, B.~S., 2010, PASP, 122, 935

\bibitem[{{Enoch} et~al.(2010){Enoch}, {Collier Cameron}, {Parley}, \&
  {Hebb}}]{Enoch+10aa}
{Enoch}, B., {Collier Cameron}, A., {Parley}, N.~R., {Hebb}, L., 2010, A\&A,
  516, A33

\bibitem[{{Enoch} et~al.(2012){Enoch}, {Collier Cameron}, \&
  {Horne}}]{Enoch+12aa}
{Enoch}, B., {Collier Cameron}, A., {Horne}, K., 2012, A\&A, 540, A99

\bibitem[{{Enoch} et~al.(2011)}]{Enoch+11mn}
{Enoch}, B., et~al., 2011, MNRAS, 410, 1631

\bibitem[{{Fortney} et~al.(2007){Fortney}, {Marley}, \&
  {Barnes}}]{Fortney++07apj}
{Fortney}, J.~J., {Marley}, M.~S., {Barnes}, J.~W., 2007, ApJ, 659, 1661

\bibitem[{{Gibson} et~al.(2009)}]{Gibson+09apj}
{Gibson}, N.~P., et~al., 2009, ApJ, 700, 1078

\bibitem[{{H{\'e}brard} et~al.(2013)}]{Hebrard+13aa}
{H{\'e}brard}, G., et~al., 2013, A\&A, 549, A134

\bibitem[{{Henden} et~al.(2012){Henden}, {Levine}, {Terrell}, {Smith}, \&
  {Welch}}]{Henden+12javso}
{Henden}, A.~A., {Levine}, S.~E., {Terrell}, D., {Smith}, T.~C., {Welch}, D.,
  2012, Journal of the American Association of Variable Star Observers, 40, 430

\bibitem[{{Knutson} et~al.(2014)}]{Knutson+14apj}
{Knutson}, H.~A., et~al., 2014, ApJ, 785, 126

\bibitem[{{Lillo-Box} et~al.(2014){Lillo-Box}, {Barrado}, \&
  {Bouy}}]{Lillobox++14aa}
{Lillo-Box}, J., {Barrado}, D., {Bouy}, H., 2014, A\&A, in press, {\tt
  arXiv:1405.3120}

\bibitem[{{L{\'o}pez-Morales}(2007)}]{Lopez07apj}
{L{\'o}pez-Morales}, M., 2007, ApJ, 660, 732

\bibitem[{{Mahtani} et~al.(2013)}]{Mahtani+13mn}
{Mahtani}, D.~P., et~al., 2013, MNRAS, 432, 693

\bibitem[{{Markwardt}(2009)}]{Markwardt07aspc}
{Markwardt}, C.~B., 2009, vol. 411 of \emph{Astronomical Society of the Pacific
  Conference Series}, p. 251

\bibitem[{{Maxted} et~al.(2011){Maxted}, {Koen}, \& {Smalley}}]{Maxted++11mn}
{Maxted}, P.~F.~L., {Koen}, C., {Smalley}, B., 2011, MNRAS, 418, 1039

\bibitem[{{McLaughlin}(1924)}]{McLaughlin24apj}
{McLaughlin}, D.~B., 1924, ApJ, 60, 22

\bibitem[{{Mortier} et~al.(2013){Mortier}, {Santos}, {Sousa}, {Fernandes},
  {Adibekyan}, {Delgado Mena}, {Montalto}, \& {Israelian}}]{Mortier+13aa}
{Mortier}, A., {Santos}, N.~C., {Sousa}, S.~G., {Fernandes}, J.~M.,
  {Adibekyan}, V.~Z., {Delgado Mena}, E., {Montalto}, M., {Israelian}, G.,
  2013, A\&A, 558, A106

\bibitem[{{Nelder} \& {Mead}(1965)}]{NelderMead65}
{Nelder}, J.~A., {Mead}, R., 1965, The Computer Journal, 7, 308

\bibitem[{{Nelson} \& {Davis}(1972)}]{NelsonDavis72apj}
{Nelson}, B., {Davis}, W.~D., 1972, ApJ, 174, 617

\bibitem[{{Nikolov} et~al.(2013){Nikolov}, {Chen}, {Fortney}, {Mancini},
  {Southworth}, {van Boekel}, \& {Henning}}]{Nikolov+13aa}
{Nikolov}, N., {Chen}, G., {Fortney}, J., {Mancini}, L., {Southworth}, J., {van
  Boekel}, R., {Henning}, T., 2013, A\&A, 553, A26

\bibitem[{{Pietrinferni} et~al.(2004){Pietrinferni}, {Cassisi}, {Salaris}, \&
  {Castelli}}]{Pietrinferni+04apj}
{Pietrinferni}, A., {Cassisi}, S., {Salaris}, M., {Castelli}, F., 2004, ApJ,
  612, 168

\bibitem[{{Poddan{\'y}} et~al.(2010){Poddan{\'y}}, {Br{\'a}t}, \&
  {Pejcha}}]{Poddany++10newa}
{Poddan{\'y}}, S., {Br{\'a}t}, L., {Pejcha}, O., 2010, New Astronomy, 15, 297

\bibitem[{{Pollacco} et~al.(2006)}]{Pollacco+06pasp}
{Pollacco}, D.~L., et~al., 2006, PASP, 118, 1407

\bibitem[{{Popper} \& {Etzel}(1981)}]{PopperEtzel81aj}
{Popper}, D.~M., {Etzel}, P.~B., 1981, AJ, 86, 102

\bibitem[{{Rossiter}(1924)}]{Rossiter24apj}
{Rossiter}, R.~A., 1924, ApJ, 60, 15

\bibitem[{{Sada} et~al.(2012)}]{Sada+12pasp}
{Sada}, P.~V., et~al., 2012, PASP, 124, 212

\bibitem[{{Simpson} et~al.(2011)}]{Simpson+11mn}
{Simpson}, E.~K., et~al., 2011, MNRAS, 414, 3023

\bibitem[{{Skottfelt} et~al.(2013)}]{Skottfelt+13aa}
{Skottfelt}, J., et~al., 2013, A\&A, 553, A111

\bibitem[{{Skrutskie} et~al.(2006)}]{Skrutskie+06aj}
{Skrutskie}, M.~F., et~al., 2006, AJ, 131, 1163

\bibitem[{{Smalley} et~al.(2010)}]{Smalley+10aa}
{Smalley}, B., et~al., 2010, A\&A, 520, A56

\bibitem[{{Smith} et~al.(2012)}]{Smith+12aa}
{Smith}, A.~M.~S., et~al., 2012, A\&A, 545, A93

\bibitem[{{Southworth}(2008)}]{Me08mn}
{Southworth}, J., 2008, MNRAS, 386, 1644

\bibitem[{{Southworth}(2009)}]{Me09mn}
{Southworth}, J., 2009, MNRAS, 394, 272

\bibitem[{{Southworth}(2012)}]{Me12mn}
{Southworth}, J., 2012, MNRAS, 426, 1291

\bibitem[{{Southworth} et~al.(2004){Southworth}, {Maxted}, \&
  {Smalley}}]{Me++04mn}
{Southworth}, J., {Maxted}, P.~F.~L., {Smalley}, B., 2004, MNRAS, 349, 547

\bibitem[{{Southworth} et~al.(2012){Southworth}, {Mancini}, {Maxted}, {Bruni},
  {Tregloan-Reed}, {Barbieri}, {Ruocco}, \& {Wheatley}}]{Me+12mn2}
{Southworth}, J., {Mancini}, L., {Maxted}, P.~F.~L., {Bruni}, I.,
  {Tregloan-Reed}, J., {Barbieri}, M., {Ruocco}, N., {Wheatley}, P.~J., 2012,
  MNRAS, 422, 3099

\bibitem[{{Southworth} et~al.(2009{\natexlab{a}})}]{Me+09mn}
{Southworth}, J., et~al., 2009{\natexlab{a}}, MNRAS, 396, 1023

\bibitem[{{Southworth} et~al.(2009{\natexlab{b}})}]{Me+09apj}
{Southworth}, J., et~al., 2009{\natexlab{b}}, ApJ, 707, 167

\bibitem[{{Southworth} et~al.(2010)}]{Me+10mn}
{Southworth}, J., et~al., 2010, MNRAS, 408, 1680

\bibitem[{{Southworth} et~al.(2013)}]{Me+13mn}
{Southworth}, J., et~al., 2013, MNRAS, 434, 1300

\bibitem[{{Stetson}(1987)}]{Stetson87pasp}
{Stetson}, P.~B., 1987, PASP, 99, 191

\bibitem[{{Street} et~al.(2010)}]{Street+10apj}
{Street}, R.~A., et~al., 2010, ApJ, 720, 337

\bibitem[{{Torres} et~al.(2010){Torres}, {Andersen}, \&
  {Gim{\'e}nez}}]{Torres++10aarv}
{Torres}, G., {Andersen}, J., {Gim{\'e}nez}, A., 2010, A\&ARv, 18, 67

\bibitem[{{Torres} et~al.(2012){Torres}, {Fischer}, {Sozzetti}, {Buchhave},
  {Winn}, {Holman}, \& {Carter}}]{Torres+12apj}
{Torres}, G., {Fischer}, D.~A., {Sozzetti}, A., {Buchhave}, L.~A., {Winn},
  J.~N., {Holman}, M.~J., {Carter}, J.~A., 2012, ApJ, 757, 161

\bibitem[{{VandenBerg} et~al.(2006){VandenBerg}, {Bergbusch}, \&
  {Dowler}}]{Vandenberg++06apjs}
{VandenBerg}, D.~A., {Bergbusch}, P.~A., {Dowler}, P.~D., 2006, ApJS, 162, 375

\bibitem[{{Zacharias} et~al.(2004){Zacharias}, {Monet}, {Levine}, {Urban},
  {Gaume}, \& {Wycoff}}]{Zacharias+05aas}
{Zacharias}, N., {Monet}, D.~G., {Levine}, S.~E., {Urban}, S.~E., {Gaume}, R.,
  {Wycoff}, G.~L., 2004, in American Astronomical Society Meeting Abstracts,
  vol.~36, p. 1418

\end{thebibliography}

\end{document}